\documentclass[manuscript,screen]{acmart}
\usepackage{shortcuts}

\usepackage{tikz}
\usepackage{amsmath}
\usepackage{xspace}
\usepackage{subcaption}	
\usepackage{enumitem}
\usepackage{booktabs}
\usepackage{adjustbox}	
\usepackage{graphicx}	
\usepackage{multirow}	
\usepackage{hyperref}
\usepackage{balance}
\usepackage{stfloats}
\usepackage[utf8]{inputenc}
\usepackage{threeparttable}
\usepackage{framed}
\usepackage{listings}
\usepackage{microtype}
\usepackage{diagbox}
\usepackage{url}
\usepackage{makecell}
\usepackage{threeparttable}
\usepackage{color}
\usepackage{xcolor}
\usepackage{colortbl}
\usepackage{mathtools}
\usepackage{nccmath}
\usepackage{algorithm}
\usepackage{algorithmic}

\usepackage{wasysym}
\newcommand{\ymark}{\CIRCLE}
\newcommand{\nmark}{\Circle}
\newcommand{\halfmark}{\LEFTcircle}

\usepackage{pifont}
\newcommand{\cmark}{\ding{52}} %
\newcommand{\xmark}{\ding{56}} %

\usepackage{tcolorbox}
\newtcolorbox{conclusionbox}{colback=gray!8,colframe=black,width=\linewidth,arc=0.6mm, boxrule=0.4pt, left=1mm,right=1mm,top=0.5mm,bottom=0.5mm,before skip=2mm,after skip=0mm}

\definecolor{patriarch}{rgb}{0.5, 0.0, 0.5}
\definecolor{acmblue}{cmyk}{1, 0.1, 0, 0.1}
\definecolor{acmdarkblue}{cmyk}{1, 0.2, 0, 0.15}
\hypersetup{
  colorlinks,
  citecolor=patriarch,
  linkcolor=patriarch,
  urlcolor=acmdarkblue
}

\setcopyright{acmlicensed}
\copyrightyear{2026}
\acmYear{2026}
\acmDOI{XXXXXXX.XXXXXXX}
\acmConference[Conference acronym 'XX]{Make sure to enter the correct
  conference title from your rights confirmation email}{June 03--05,
  2026}{Woodstock, NY}
\acmISBN{978-1-4503-XXXX-X/2026/06}

\begin{document}

\title{Unraveling the Key of Machine Learning-based Android Malware Detection}

\author{Jiahao Liu}
\affiliation{%
  \institution{National University of Singapore}
  \city{Singapore}
  \country{Singapore}}
\email{ljiahaomail@gmail.com}
\email{jiahao99@comp.nus.edu.sg} 
\author{Jun Zeng}
\authornote{For correspondence, please contact Jiahao Liu and Jun Zeng.}
\affiliation{%
  \institution{National University of Singapore}
  \city{Singapore}
  \country{Singapore}}
\email{junzeng@comp.nus.edu.sg}
\author{Fabio Pierazzi}
\affiliation{
  \institution{University College London}
  \city{London}
  \country{United Kingdom}}
\email{f.pierazzi@ucl.ac.uk}
\author{Ziqi Yang}
\affiliation{%
  \institution{The State Key Laboratory of Blockchain and Data Security, Zhejiang University}
  \city{Hangzhou}
  \country{China}}
\affiliation{
  \institution{Hangzhou High-Tech Zone (Binjiang) Institute of Blockchain and Data Security}
  \city{Hangzhou}
  \country{China}}
\email{yangziqi@zju.edu.cn}
\author{Lorenzo Cavallaro}
\affiliation{
  \institution{University College London}
  \city{London}
  \country{United Kingdom}}
\email{l.cavallaro@ucl.ac.uk}
\author{Zhenkai Liang}
\affiliation{%
  \institution{National University of Singapore}
  \city{Singapore}
  \country{Singapore}}
\email{liangzk@comp.nus.edu.sg}

\renewcommand{\shortauthors}{Jiahao Liu et al.}
\begin{abstract}
With the rapid advancement of machine learning (ML), ML-based Android malware detection has gained significant popularity due to its ability to automatically learn malicious patterns from Android apps.
However, the lack of an in-depth and systematic analysis of existing research makes it difficult to obtain a holistic understanding of the state of the art in this field.
In this work, we present the most comprehensive investigation to date of ML-based Android malware detection systems, combining both empirical and quantitative analyses.
We first organize prior work into a unified taxonomy based on Android app representations and the ML modeling pipeline.
Building on this taxonomy, we design a general-purpose framework for ML-based Android malware detection and re-implement 12 representative approaches from three research communities—software engineering, security, and machine learning.
Using this framework, we conduct a large-scale evaluation across three key dimensions: detection effectiveness, robustness to real-world challenges, and efficiency.
Despite extensive research efforts and encouraging results, our findings reveal that existing learning-based Android malware detectors still face significant challenges, including vulnerability to malware evolution and susceptibility to adversarial attacks.
We attribute these limitations to the detectors' ability to capture and leverage malware semantics, defined as semantic information that characterizes malicious behaviors derived from APK features.
Finally, we summarize our key insights and provide actionable recommendations to guide future research in this domain.

\end{abstract}


\maketitle
\section{Introduction}
\label{sec:intro}

Over the past decade, ML-based Android malware detection has attracted increasing attention from various research communities, such as software engineering, security, and machine learning~\cite{qiu2020survey,liu2022deep,wu2019malscan,wu2021homdroid,mclaughlin2017deep,he2022msdroid,xu2018deeprefiner,aafer2013droidapiminer,arp2014drebin}.
Much attention has been given to exploring various combinations of APK features and ML models.
The trend is primarily led by advancements in ML models (\textit{e.g.}, graph neural network~\cite{he2022msdroid}) and analogies drawn from other well-studied fields (\textit{e.g.}, social network~\cite{wu2019malscan}).
Most existing approaches report high F1 scores (up to 0.98)~\cite{wu2019malscan,kim2018multimodal}.
Such promising results motivate us to ask a number of important research questions \tosem{regarding the current state of ML-based Android malware detection.}
For example,
How do existing approaches represent and incorporate diverse APK features into ML models? 
How do different detection methods compare when evaluated under the same datasets, metrics, and toolchains?
Are current ML-based approaches sufficient to meet real-world deployment requirements?
Does employing more powerful ML models necessarily lead to better detection performance? Does model selection materially affect detection effectiveness when the feature set is fixed?
Does incorporating more features to describe app behavior always improve performance?
Finally, is there a positive correlation between detection effectiveness and computational efficiency?

Existing studies~\cite{faruki2014android,qiu2020survey,liu2022deep,gao2024comprehensive} provide a good starting point to understand the landscape of ML-based Android malware detection.
However, most are either theoretical reviews or limited in experimental scope and scalability, and thus fall short of a systematic, end-to-end view of the field.
For example, Qiu et al.~\cite{qiu2020survey} organize the landscape by model family (\textit{e.g.,} CNN and RNN) and examine how these models are applied in Android malware detection.
To our knowledge, no prior work provides an empirical and quantitative real-world analysis that jointly considers effectiveness, robustness, and efficiency.
Additionally, there is no publicly available framework that standardizes the implementation and evaluation of Android malware detectors, enabling reproducible research and supporting future development.

Accordingly, we conduct an empirical, quantitative study to answer these questions and clarify the current state of ML-based Android malware detection.
Given the scale of the ecosystem --- Google Play's catalog reached 4.2 million apps in 2025, a 6.3\% increase over 2024~\cite{appnumber} --- we prioritize methods that scale to large datasets and are practical for real-world deployment.
Specifically, we implement and evaluate representative approaches to characterize both progress and remaining gaps.
While informative, this exercise also reveals several challenges to fair comparisons as follows.

\begin{itemize}[leftmargin=10pt, itemsep=2pt]
  \item \textit{Unfair Comparisons.}
  Previous approaches are usually evaluated on datasets of different sizes, goodware-to-malware ratios, and training-to-testing ratios~\cite{arp2014drebin,hou2017hindroid,li2021robust}.
  Moreover, they report outcomes using diverse metrics (\textit{e.g.,} F1-score, Accuracy, and False Positive Rate), making it unclear which method performs better under specific settings.
  In addition, they often rely on different toolchains (\textit{e.g.,} Androguard~\cite{androguard} and APKTool~\cite{apkt}) to develop detectors, which introduce ambiguity --- whether the promising results stem from the novelty of the approach or not~\cite{vasilache2018tensor}.

  \item \textit{Unrealistic Evaluations.} 
  Android malware detectors face a rapid threat landscape~\cite{barbero2022transcending}, with malware continually evolving to evade detection.
  The growing usage of obfuscation also makes malware harder to detect as its malicious intentions are hidden~\cite{aonzo2020obfuscapk}.
  Additionally, ML models can be tricked by intentionally perturbed inputs~\cite{pierazzi2020problemspace,carlini2017towards,suya2020hybrid,grosse2017adversarial}.
  Although the impacts of some of these scenarios have been studied~\cite{jordaney2017transcend, barbero2022transcending,pendlebury2019tesseract,ceschin2020machine}, the lack of a comprehensive assessment creates a gap in our understanding of the main challenges that hinder the adoption of Android malware detection in real-world scenarios.

  \item \textit{Unclear Computational Costs.}
  With the exponential growth of apps in sizes and complexities, feature extraction gradually becomes notably time-consuming --- for example, it can take up 30 minutes to gather API paths from one app~\cite{xu2020sdac}.
  Furthermore, as ML models evolve in complexity, they demand greater computational power to achieve state-of-the-art results.
  Unfortunately, it remains uncertain what prices to pay to gain the desired results.
\end{itemize}

To address these challenges, we design \codename, a general-purpose framework that streamlines the implementation and evaluation of ML-based Android malware detection systems.
\codename adopts a modular and configurable architecture, enabling both the rapid development of new detection techniques and the flexible construction of diverse evaluation scenarios.
Specifically, we begin with decomposing the detection pipeline into three core phases: APK characterization, feature representation, and ML modeling.
Guided by this taxonomy, we conduct an empirical analysis of prior work to examine how existing approaches detect Android malware.

Next, we apply our framework to 12 representative approaches spanning three research communities: software engineering~\cite{wu2019malscan,wu2021homdroid,wu2021android}, security~\cite{arp2014drebin,mariconti2016mamadroid,mclaughlin2017deep,xu2018deeprefiner,kim2018multimodal,xu2020sdac,he2022msdroid}, and machine learning~\cite{hou2017hindroid,li2021robust}.
To ensure fairness and reproducibility, we standardize shared tasks (\textit{e.g.,} feature extraction) across different methods by using the same toolchain (\textit{i.e.,} Androguard~\cite{androguard}), eliminating discrepancies caused by heterogeneous tool support.
Furthermore, to facilitate the development of new detectors, \codename is implemented as a modular and configurable system, allowing components (\textit{e.g.,} neural networks) to be easily replaced or customized.

For a comprehensive assessment, we randomly sample 221,310 apps spanning ten years (2011-2020) from AndroZoo~\cite{allix2016androzoo}, following prior studies~\cite{pendlebury2019tesseract,chen2023continuous}.
AndroZoo is a continuously updated public repository of Android applications, and this dataset serves as our primary benchmark.
We set the malware ratio to 10\% to reflect realistic deployment conditions~\cite{pendlebury2019tesseract}.
To systematically investigate the state of ML-based Android malware detection, we evaluate the selected approaches using commonly adopted metrics (\textit{e.g.}, F1-score and accuracy), focusing on multiple dimensions: detection effectiveness and efficiency, robustness to app evolution and obfuscation, and resilience against adversarial attacks.
To further validate the generalizability of our findings, we construct an additional dataset consisting of 7,911 apps collected between 2021 and 2024, including 7,120 benign apps from AndroZoo and 791 malicious apps from VirusTotal and AndroZoo.
We evaluate the effectiveness of the selected approaches on this supplementary dataset to examine whether our key observations remain valid on more recent data.

\noindent \textbf{Findings and Recommendations.} Through our empirical and quantitative analysis, we present a holistic view of the state of the art in ML-based Android malware detection.
Our results show that APK characterization and ML modeling remain central themes in the literature and continue to serve as the foundation for building effective detectors.
Under identical experimental settings, many recent approaches achieve comparable performance.
However, their effectiveness still degrades in challenging scenarios, such as those involving limited training data or adversarial attacks.
Our analysis indicates that the key factor underlying detection effectiveness is the ability to extract semantic information that characterizes malicious behaviors from APK features, which we refer to as malware semantics.
From the feature-design perspective, we observe that naively incorporating additional features does not necessarily improve performance; instead, irrelevant or weakly related features may dilute meaningful semantic signals and even harm detection accuracy.
From the modeling perspective, we find that employing more powerful ML models alone does not guarantee better detection outcomes, particularly when the input features fail to adequately capture app behavior.
Furthermore, given specific feature sets (e.g., API calls and permissions), ensemble-based models and deep learning models consistently outperform some traditional classifiers, suggesting that these models constitute strong and practical baselines for future detector design.
We also observe that increased computational overhead is not a reliable indicator of improved detection capability, highlighting the need to carefully balance model complexity and performance.
Overall, our findings suggest that future research should place greater emphasis on designing robust and practical detectors for real-world deployment, with careful consideration of both effectiveness and efficiency.
Additional discussions and insights are provided in Section~\ref{sec:findings}.

We emphasize that the design of effective ML-based Android malware detection solutions should be driven by the extraction and integration of malware semantics --- that is, semantic representations of malicious app behaviors derived from APK features.
Defining, modeling, and quantifying such semantics remains an open research challenge, yet it is critical for advancing the effectiveness and robustness of ML-based detection methods.
We hope that our findings help inform and guide future research in this area, including --- but not limited to --- feature selection, model design, and the efficient allocation of computational resources.

In summary, we make the following contributions:
\begin{itemize}[leftmargin=10pt, topsep=0pt, itemsep=2pt]
  \item
  We conduct a thorough systematic investigation of ML-based Android malware detection using empirical and qualitative methods, drawing a holistic picture of the field.
  \item
  We design a general-purpose framework, \codename, to facilitate the implementation and evaluation of various Android malware detectors.
  For comparison in realistic settings, we collect the largest dataset to date, both in size and temporal coverage. To promote reproducibility and future research, we release our framework and dataset at \url{https://github.com/ljiahao/FrameDroid}.
  \item 
  We offer a comprehensive comparative analysis of 12 representative approaches using \codename, focusing on assessing their effectiveness, robustness, and efficiency. We point out that the key to enhancing ML-based Android malware detection is incorporating the malware semantics derived from APK features.
\end{itemize}

\section{Learning-based Android Malware Detection}
\label{sec:systematic_investigation}
Android malware detection involves two steps: characterizing APKs and identifying malicious patterns.
Recent trends have seen a shift towards using static feature extraction for APK profiling due to its efficiency and scalability~\cite{qiu2020survey,liu2022deep}.
For malicious pattern identification, ML models have become increasingly popular since they can automatically learn patterns from features~\cite{wu2019malscan,mclaughlin2017deep,li2021robust}.
To validate the trend towards ML-based Android malware detection, we conduct a systematic investigation of the literature, which can be found in Section~\ref{sec:discussion}.

It is worth noting that, in an effort to better characterize the landscape of ML-based Android malware detection, a number of studies~\cite{souri2018state,qiu2020survey,liu2022deep,faruki2014android,gao2024comprehensive} have reviewed existing detection approaches to identify common practices and open challenges.
However, these reviews largely rely on theoretical analyses without experimental validation or adopt narrow perspectives --- for example, focusing primarily on how popular ML models are applied to this domain~\cite{qiu2020survey} or examining individual solutions in isolation.
Such limitations make it difficult to draw broader conclusions or form a holistic understanding of how the entire detection pipeline operates, including what features are commonly used and how these features are utilized by ML models.
Gao et al.~\cite{gao2024comprehensive} attempt to experimentally assess the state of ML-based Android malware detection; however, the analysis remains limited in scope and scale (\textit{e.g.,} focusing mainly on robustness) and does not account for real-world settings or end-to-end performance evaluation.
To the best of our knowledge, no existing work provides a comprehensive study that systematically investigates ML-based Android malware detection through both (i) empirical categorization and elucidate the overall detection workflow, and (ii) quantitative evaluation to assess the state of the art across multiple dimensions --- effectiveness, robustness, and efficiency --- under realistic conditions.
Moreover, there is currently no publicly available, general-purpose framework that supports the development and evaluation of ML-based Android malware detection systems.

To bridge this gap, we first present a systematic investigation of ML-based Android malware detection.
We then design and release a general-purpose framework, \codename, together with a large-scale, realistic dataset (see Section~\ref{sec:framework}), enabling the most comprehensive and wide-ranging comparative analysis to date.
Using this unified platform, we examine the current state of ML-based Android malware detection and highlight the key challenges and opportunities that shape this field.

In this section, we demonstrate the empirical analysis about how existing detectors represent and incorporate features for Android malware detection. 
Rather than analyzing each approach individually, our methodology is to deconstruct and unify the workflow of ML-based Android malware detection. We identify three common phases: APK characterization (Section~\ref{sub:apk_characterization}), feature representation (Section~\ref{sub:apk_feature_encoding}), and ML modeling (Section~\ref{sub:ml_models}).
Following this, we collate and summarize the key techniques employed in each phase by investigating existing approaches, offering a thorough overview of how ML models are applied in Android malware detection.

\subsection{APK Characterization}
\label{sub:apk_characterization}
\bulletpoint{Input from APK}
An APK is a compressed archive that contains an app's codebase, resources, and auxiliary files.
It contains several types of files and directories, including the Manifest ($\mathbb{M}$), Dex ($\mathbb{D}$), Library ($\mathbb{L}$), and Resource ($\mathbb{R}$) components~\cite{faruki2014android,apkfile}, each contributing distinctive features relevant to malware detection.
\textit{Manifest}: The manifest serves as a descriptor file that provides essential metadata about the app, including the package name, requested permissions, components, and hardware requirements.
\textit{Dex}: This includes Java classes that are compiled according to the Dalvik Executable (DEX) file standard, designed to run on the Dalvik Virtual Machine (DVM).
\textit{Library}: Native libraries appear as shared object files and offer critical low-level functionality (\textit{e.g.,} WebKit). They support and optimize the app’s execution and may expose behaviors indicative of malicious activity.
\textit{Resource}: This category includes static assets and non-compiled resources required by the app, such as images, XML layouts, and animation sequences.

\bulletpoint{APK features}
The quartet of file types described above forms the foundation of APK characterization. Researchers commonly employ static analysis techniques to extract and distill relevant features.
\jh{\textit{Manifest} and \textit{Resource} files typically follow well-defined structures, such as XML,} which makes them straightforward to parse. Features from these components can thus be efficiently extracted using regular expressions or dedicated XML parsers.
In contrast, the \textit{Dex} and \textit{Library} components consist of binary files, making their analysis substantially more involved. Feature extraction from these binaries requires reverse engineering techniques and specialized tooling.
\textit{Dex} can be disassembled by Androguard~\cite{androguard} and APKTool~\cite{apkt} into smali code, which is a more human-readable representation depicting apps' behaviors.
For the \textit{Library}, tools like Angr~\cite{angr} or IDA Pro~\cite{ida_pro} are helpful in disassembling the native libraries into assembly codes, which facilitates analysis of the native services utilized by apps.
By analyzing this assembly code, it is possible to capture critical features within native code.

\begin{figure}[t]
    \centering
    \includegraphics[width=0.68\linewidth]{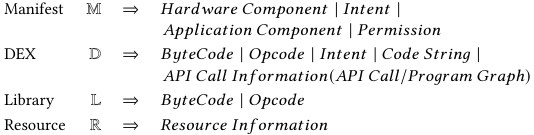}
    \vspace{-0.2cm}
    \caption{APK files and their corresponding features.}
    \label{fig:features}
    \vspace{-0.6cm}
\end{figure}

Figure~\ref{fig:features} delineates the features typically utilized and their corresponding APK file or folder in ML-based Android malware detection.
In the subsequent section, we elaborate on these features in detail.

\featurepoint{[M]}{Hardware Component}
Android apps employ certain hardware components (\textit{e.g.,} camera) to execute particular functions (\textit{e.g.,} taking photos).
The request for specific hardware components carries distinct security implications, as the utilization of hardware combinations often indicates potentially harmful behaviors~\cite{arp2014drebin}.
For instance, an app utilizing the camera and network may have the capability to monitor user activities and transmit this data to remote servers.
Recognizing the potential of this insight, several approaches~\cite{arp2014drebin,xu2018deeprefiner,kim2018multimodal,wang2019effective} have exploited these features as a heuristic for Android malware detection.

\featurepoint{[M]}{Application Component}
An APK uses four primary components, namely, Activity, Service, Content Provider, and Broadcast Receiver, to provide different entry points for the system/users.
Specifically, Activity provides interfaces for direct user engagement; Service sustains the app's background operations; Broadcast Receiver delivers system-wide events to the app, and Content Provider manages a shared set of app data.
Commonly, one malware family employs similar component names, such as \textit{SearchService} in the DroidKungFu family~\cite{kungfu}.
Inspired by this, application components are utilized to capture similar fingerprints in Android malware~\cite{xu2018deeprefiner,kim2018multimodal,wang2019effective}.

\featurepoint{[M|D]}{Intent}
As the primary ways of communication among components, intents connect various Application Components and delineate standard operations one app can perform.
They are pivotal in initiating Activities, managing Services lifecycle, and delivering broadcast information to Broadcast Receivers.
Malware often monitors these communication status to trigger malicious actions, such as activating pre-configured malicious activities~\cite{zhou2012hey}.
Approaches~\cite{li2021robust,feng2020performance,xu2016iccdetector} aim to capture these malicious behaviors by analyzing corresponding intents.

\featurepoint{[M]}{Permission}
Android employs a permission-based mechanism to regulate access to sensitive data and restricted actions.
For an app to carry out particular actions, it must obtain the requisite permissions~\cite{au2012pscout,backes2016demystifying}.
The set of permissions required by an app can thus offer insights into its intended behaviors.
Particularly, malware often demands permissions that are unnecessary for benign apps to execute malicious actions~\cite{enck2009lightweight}.
Consequently, numerous Android malware detectors~\cite{he2022msdroid,xu2018deeprefiner,arp2014drebin,kim2018multimodal,wu2021android} capture the differences to distinguish malware.

\featurepoint{[D]}{API Call Information (API Call and Program Graph)}
API Call Information, consisting of API calls and their connections, is a widely utilized feature source in Android malware detection.
Android apps make use of these API calls to access the operating system's functionality and system resources~\cite{peiravian2013machine}.
For instance, the invocation of \textit{sendTextMessage()} suggests that the app is likely to send a text message.
Furthermore, API calls are connected to form a graph, where each node signifies a method, and each edge denotes a method invocation~\cite{li2023black}, illustrating the app's structural information~\cite{wu2019malscan}.
For clarity, we differentiate API Call Information into two sub-categories: \textit{API Call}, referring to the individual API calls, and \textit{Program Graph}, denoting the relationships among these API calls.
Numerous studies~\cite{arp2014drebin,kim2018multimodal,wu2021android} detect sensitive API calls (\textit{e.g., getDeviceId()}) to estimate the probability of an app being malicious.
In contrast, other solutions~\cite{wu2019malscan,he2022msdroid,hou2017hindroid,xu2020sdac,martin2017evolving,ye2019out} venture deeper, examining the relationships between API calls to capture an app's semantics for malware detection.

\featurepoint{[D|L]}{ByteCode and Opcode}
Similar to previous research~\cite{xu2018deeprefiner,sun2016taintart,li2021palmtree,daoudi2021dexray}, we treat both the \textit{raw bytecode} and the \textit{assembly code} derived from the \textit{Dex} and \textit{Library} as ByteCode.
ByteCode contains a sequence of instructions, where each instruction consists of a single Opcode and several operands.
The Opcode denotes a specific operation; for instance, the \texttt{invoke} means a method invocation.
The operands provide additional information for the Opcode, such as the method name.
In addition, ByteCode and Opcode from the \textit{Dex} and \textit{Library} offer insights into the static execution flows of apps' Java and Native codes, providing a view of how an app work~\cite{kim2018multimodal}.
Recent research attempts to represent Bytecode and Opcode in various formats, such as image~\cite{mclaughlin2017deep,hsien2018r2,amin2022android,xiao2019image} and text~\cite{xu2018deeprefiner,yan2018lstm,karbab2021petadroid}, to capture apps' semantics.

\featurepoint{[D]}{Code String}
Apps often embed key information like URLs and IP addresses as string values within their codebase.
These strings can be traced in the assembly code, tagged either as \texttt{const-string} or \texttt{const-string/jumbo}. 
Such strings can provide crucial clues about potential malicious activities~\cite{arp2014drebin}.
For instance, malware sets up network sockets to communicate with remote servers, using the string \texttt{socket} in the codebase.
A number of works~\cite{arp2014drebin,kim2018multimodal,zhu2019transparent} have used code strings to identify potential illegal operations.

\featurepoint{[R]}{Resource Information}
Apps utilize resources to house traditional files and static elements, such as bitmaps and animation instructions.
These resources are generally decoupled from the application codebase for ease of maintenance.
Attackers sometimes embed malicious code within resource files, like image files, as a tactic to evade detection.
Approaches like DeepRefiner~\cite{xu2018deeprefiner} consider resource information as a crucial feature source for malware detection.

\begin{figure}[t]
    \centering
    \includegraphics[width=0.68\linewidth]{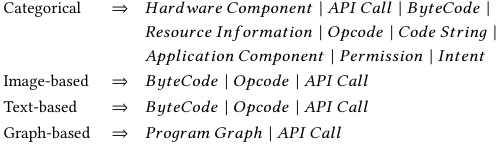}
    \vspace{-0.1cm}
    \caption{The relationships between Feature Representations and widely used APK Features.}
    \label{fig:representation}
    \vspace{-0.5cm}
\end{figure}

\subsection{Feature Representation}
\label{sub:apk_feature_encoding}
APK characterizations provide multi-perspective views of an app's behavior.
Before these features can be fed into ML models, they must be encoded into representations that the models can interpret.
Broadly, the encoding process falls into four categories: categorical, image-based, text-based, and graph-based.
\jh{Figure~\ref{fig:representation} illustrates the relationships between these encoding strategies and the corresponding APK features.}
In the following section, we discuss each encoding strategy in detail.

\bulletpoint{Categorical encoding}
Features like hardware components, intents, and code strings are often viewed as categorical data and can be easily transformed into numerical values.
A widely-encoding strategy is to convert these features into a binary vector, where each position indicates whether a specific feature exists or not~\cite{arp2014drebin,hou2017hindroid,li2021robust,wu2021android,yuan2014droid,li2019adversarial,feng2019mobidroid}.
On the other hand, another line of research~\cite{kim2018multimodal} calculates the frequency of each feature to obtain a vector of numerical values, reflecting the importance of each feature.

\bulletpoint{Image-based encoding}
Representing specific features as images and leveraging image processing techniques is a well-established approach in malware detection.
Specifically, bytecode and opcode are often visualized as images to describe apps' behaviors~\cite{mclaughlin2017deep,daoudi2021dexray,hsien2018r2,xiao2019image,ding2020android}.
Notably, \textsc{DexRay}~\cite{daoudi2021dexray} transforms the app's bytecode into grey-scale vector images, wherein each pixel corresponds to a distinct byte.
In a similar vein, some other features are also mapped to images to depict the APK's characteristics.
For instance, Zegzhda et al.~\cite{zegzhda2018applying} combine API calls with protection levels as an RGB image.

\bulletpoint{Text-based encoding}
Text-based encoding is also a widely used strategy in Android malware detection.
Many existing methods~\cite{xu2018deeprefiner,xu2020sdac,karbab2021petadroid,karbab2018maldozer,sun2019scalable} have approached APK features from a textual perspective, employing natural language processing (NLP) techniques to amplify detection capability.
For example, by considering API calls as \texttt{words} and their sequences as \texttt{sentences}, methods presented in~\cite{xu2020sdac,karbab2021petadroid,karbab2018maldozer} utilize word embedding techniques, such as Word2Vec~\cite{mikolov2013distributed}, to extract semantic information included in the API calls.
Separately, Sun et al.~\cite{sun2019scalable} treat API calls and permissions as sparse data, applying Doc2Vec~\cite{le2014distributed} to derive their vector representations.

\bulletpoint{Graph-based encoding}
Recently, graph structure has been widely adopted to represent apps' semantics~\cite{he2022msdroid,li2023black}.
One direction is to leverage program graphs to model APK behaviors~\cite{wu2019malscan,wu2021homdroid,he2022msdroid,pektacs2020deep,pektacs2020learning}.
Another avenue aims to build API-based feature graphs, drawing insights from API calls and their meta-relationships. 
An example is to identify whether two API calls are in the same block, thereby capturing the app's intended operations~\cite{hou2017hindroid,huang2019deep}.
When these features are represented as graphs, graph-based techniques (\textit{e.g.,} Graph2Vec~\cite{narayanan2017graph2vec}, social network~\cite{wu2019malscan}) are employed to extract apps' structural information.

\subsection{Machine Learning Modeling}
\label{sub:ml_models}
After encoding these features as numerical vectors, machine learning models are leveraged to identify malicious patterns.
Following previous studies~\cite{qiu2020survey,zhao2021impact}, we categorize the models employed in Android malware detection into two main categories: traditional machine learning (TML) and deep learning (DL) models.
TML models, such as linear regression or decision trees, typically exhibit simpler structures that can explicitly model the relationship between input and output. 
As such, these TML models often require domain knowledge to extract features from input data.
In contrast, DL models are characterized by their multiple layers of neurons, enabling them to capture complex non-linear mappings from input to output~\cite{zhao2018deepsim}.
This capability means that DL models are less reliant on domain knowledge during the feature extraction.
In this section, we provide a concise introduction to the widely used models.

\bulletpoint{TML models}
Given APK features, TML models are commonly employed to discern patterns from them.
The Support Vector Machine (SVM) can find one hyperplane that separates the high-dimensional data points with varying labels.
This capability has made it a popular choice to detect Android malware~\cite{arp2014drebin,hou2017hindroid,xu2020sdac,sun2016sigpid,garcia2018lightweight}.
K-Nearest Neighbors (KNN) has also been applied in Android malware detection~\cite{wu2019malscan,wu2021homdroid,aafer2013droidapiminer,wu2016effective}.
This algorithm identifies the nearest neighbors of a given sample and subsequently classifies the sample based on the majority label of its neighbors.
Additionally, as an ensemble-based learning method, Random Forest (RF) creates a forest of decision trees, each trained on a random subset of the data.
This approach capitalizes on the strength of multiple decision trees, making the model more robust and accurate than individual trees.
Such advantages have led to its significant application in malware detectors~\cite{mariconti2016mamadroid,zhu2018hemd}.

\bulletpoint{DL models}
DL models have exhibited strong capability in modeling malware behaviors. 
This part provides an insight into the widely used DL models tailored for Android malware detection.
As a basic feed-forward neural network, Multi-Layer Perceptron (MLP), composed of several layers of neurons, \jh{has shown significant effectiveness in detecting Android malware~\cite{kim2018multimodal,li2021robust,martin2017evolving,sun2019scalable,zhu2019fsnet,chen2020denas}.}
The Recurrent Neural Network (RNN)~\cite{medsker2001recurrent} is a type of neural architecture that can capture the sequential information of input data.
By representing APK features (\textit{e.g.}, API calls and bytecode) as sequences, existing solutions~\cite{xu2018deeprefiner,yan2018lstm,vinayakumar2017deep} leverage RNN to explore the temporal dependencies embedded in these features.
The Convolutional Neural Network (CNN)~\cite{he2017mask} is equipped with multiple convolutional and pooling layers, enabling it to recognize contextual information derived from low-level features~\cite{girshick2015fast}.
This intrinsic capability makes it especially effective in capturing patterns from image-oriented data.
Reflecting its efficacy, CNN has been extensively employed to extract malicious patterns from image-based features in Android malware detection~\cite{mclaughlin2017deep,hsien2018r2,feng2019mobidroid,karbab2018maldozer,huang2019deep,he2019malware}.
Given the graph representation of APK features (\textit{e.g.,} program graphs), Graph Neural Network (GNN)~\cite{kipf2016semi} can facilitate malware detection~\cite{he2022msdroid,mariconti2016mamadroid,gao2021gdroid,fan2021heterogeneous}.
This is because GNN can effectively propagate and aggregate node information along graph edges, thereby capturing the structural information of apps.
Utilizing a process of encoding and subsequently decoding input features, Autoencoders (AE) have the capacity to generate refined data representations, which makes it a popular choice in detecting malware~\cite{li2021robust,yang2021cade,zhu2021hybrid}.

In addition, existing research~\cite{su2016deep,wang2020sedroid,amin2022android} also tries to explore the potential of other DL models, like generative adversarial network (GAN)~\cite{goodfellow2014generative} and deep belief network (DBN)~\cite{hinton2009deep}.
For instance, Amin et al.~\cite{amin2022android} employ the dual-network structure of GAN --- one generates malware samples and the other works to distinguish these samples --- to enhance the malware detection capability.

\begin{center}
    \begin{conclusionbox}
    \textit{Finding:} 
    \jh{At the core of static and ML-based Android malware detection lies the ability to accurately \textit{profile app behaviors} and to \textit{select and leverage appropriate ML models} that can effectively uncover malicious patterns. Achieving this requires not only expressive feature representations but also a careful alignment between behavioral semantics and model capabilities.}
    \end{conclusionbox}
\end{center}

\section{\codename}
\label{sec:framework}

\begin{figure}[t]
    \centering
    \includegraphics[width=0.84\linewidth]{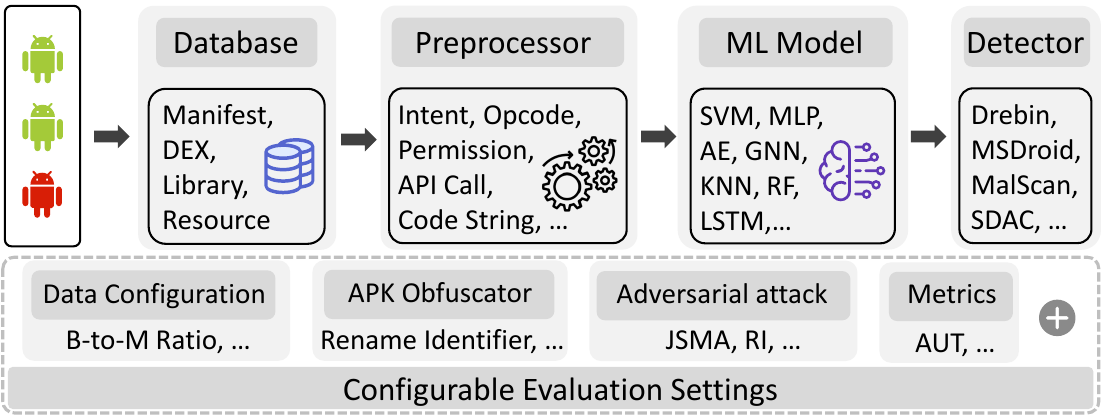}
    \vspace{-0.1cm}
    \caption{The overall architecture of \codename.}
    \label{fig:framework}
    \vspace{-0.4cm}
\end{figure}

To support experimental reproducibility and fair comparison, we propose a general-purpose framework, \codename, guided by the detection workflow discussed in Section~\ref{sec:systematic_investigation}, to facilitate the development and evaluation of new Android malware detectors.
Figure~\ref{fig:framework} illustrates the overall architecture and workflow of \codename.
The process begins with a collection of apps, from which a set of features is extracted and stored in a {\em feature database}.
These features are then processed by a {\em preprocessor}, which transforms them into numerical vectors.
A selected {\em ML model} is subsequently trained and evaluated to detect malicious applications.
Furthermore, \codename provides configurable experimental settings to support diverse real-world scenarios, such as varying goodware-to-malware ratios, different training data sizes, and robustness assessments against adversarial attacks.

This modular design of \codename simplifies the development of ML-based Android malware detectors. Three key modules, \textit{feature database}, \textit{preprocessor}, and \textit{ML model}, captures the main aspects of designing ML solutions for Android malware detection.
The \textit{feature database} organizes extracted features categorically according to their sources, as discussed in Section~\ref{sub:apk_characterization}, including Manifest, Dex, Library, and Resource. 
For example, the Manifest category contains features extracted from the AndroidManifest.xml file, such as permissions, intents, and application components.
Further details about the feature database are provided in Appendix~\ref{sub:feature_db}.
Notably, all features are stored in a manner consistent with our APK characterization taxonomy.
Each directory corresponds to a specific feature source, and the features of a given APK are distributed accordingly --- for instance, Hardware Components and Application Components under Manifest; API Calls, Opcodes, and Program Graphs under Dex; Native Functions under Library; and Image Resources under Resource.
Such structured storage enables developers to easily query and extract the required features without re-extracting them from the original APKs, greatly accelerating the development of new detectors.
In addition, the modular design allows new feature types to be seamlessly integrated into the framework.

The \textit{preprocessor} retrieves and encodes features before they are fed into the ML model.
This module can be customized to support different detectors, providing flexibility in feature selection and usage.
Leveraging the structured organization of the \textit{feature database}, users can easily select and combine features from multiple sources to construct customized feature vectors.
For instance, when designing a new detector, one can quickly retrieve a subset of features at the APK-characterization level using the \textit{preprocessor}.
To replicate a detector that relies on API calls and permissions, the \textit{preprocessor} can be configured to extract these features from the Dex and Manifest categories, respectively, and encode them into a unified feature vector.

The \textit{ML model} module integrates a wide range of commonly used machine learning models, such as RF, SVM, KNN, MLP, LSTM, CNN, GNN, and AE.
Each model offers user-friendly interfaces for training, testing, and evaluation, and the module supports easy customization and extension --- for example, adding new neural network architectures.
To develop a new detector, users only need to customize the \textit{preprocessor} to select the required features from the \textit{feature database} and feed them into the chosen model within the \textit{ML model} module.
If certain features or models are not yet supported, the framework can be readily extended by adding new feature types to the \textit{feature database} and incorporating additional learning models into the ML module.

\codename provides configurable experimental settings to support different evaluation scenarios.
Parameters, such as training sample size, goodware-to-malware ratios, and temporal distribution of samples, are all adjustable.
The framework offers fine-grained control over these parameters and supports checkpointing during both training and evaluation.
This flexibility facilitates app selection and scenario construction, ensuring comprehensive and accurate evaluations.
For example, users can configure the goodware-to-malware ratio to assess model performance under varying malware prevalence.
They can also adjust the temporal distribution of apps to simulate malware evolution, which is essential for evaluating model robustness.
Furthermore, with flexible training-sample configuration and organized checkpoint management, \codename enables incremental training, allowing users to incorporate new training samples and continue training from existing models.

\codename also integrates an APK obfuscator~\cite{aonzo2020obfuscapk}, which can produce various types of obfuscated samples, \textit{e.g.}, identifier renaming, resource encryption, code modification, and invocation reflection.
Obfuscation is applied to APKs before feature extraction to generate their obfuscated variants.
The different obfuscation types can be applied as independent operations: one can first apply one type of obfuscation and then apply another to an already obfuscated APK. 
This capability is essential for evaluating the model's resilience to obfuscated malware.
Moreover, this framework incorporates an adversarial sample generation mechanism from AndroidHIV~\cite{chen2019android}, enabling the assessment of model resilience to adversarial attacks.
Specifically, adversarial samples are generated by perturbing the feature vectors of original samples, and these perturbed samples are then used to evaluate the robustness of the trained models.
The perturbation process ensures that the functionality of the original apps is preserved.
Further details can be found in AndroidHIV~\cite{chen2019android}.
Such flexibility and adaptability make \codename a versatile tool for developing and evaluating ML-based Android malware detectors.

\bulletpoint{\codename Implementation}
\codename is developed in 17K lines of Python code.
To ensure consistency and mitigate biases from different feature extraction toolchains and learning frameworks, we adopt a standardized setup across all Android malware detectors.
For feature extraction, we use Androguard~\cite{androguard} to disassemble APK files and derive features such as permissions, intents, and program graphs. 
LibRadar~\cite{libradar} helps in identifying third-party libraries within the applications, while Angr~\cite{angr} is used for analyzing native libraries and capturing essential features like opcodes and API calls. 
When it comes to machine learning models, the scikit-learn library is our choice for traditional ML algorithms like SVM, KNN, DT, and RF.
On the other hand, for deep learning architectures such as CNN, GNN, and AE, we resort to Pytorch~\cite{pytorch}.
This uniform approach ensures a balanced evaluation, concentrating purely on the uniqueness and performance of each method.

\bulletpoint{Potential usage and availability of \codename}
As discussed above, \codename is designed with a modular and flexible architecture, making it easy to extend and customize to meet diverse research and evaluation requirements.
The framework also incorporates a comprehensive app dataset that spans a wide range of Android applications across different time periods.
Researchers and practitioners can readily leverage \codename to develop new ML-based Android malware detectors, evaluate their performance under diverse scenarios, and compare them with existing approaches in a fair and consistent manner.
For example, when designing a new detector based on API calls and permissions, users can easily retrieve these features from the \textit{feature database} and feed them into their proposed model.
Likewise, existing detectors can be readily reproduced by appropriately configuring the \textit{preprocessor} and selecting the corresponding models from the \textit{ML model} module, enabling systematic and reproducible comparisons.
It has already been adopted by several academic and industrial organizations across multiple countries, including the United States, Singapore, and China, for the development and evaluation of ML-based Android malware detectors.
This level of adoption demonstrates the practicality and effectiveness of \codename in supporting malware analysis research.

\vspace{-0.1cm}
\section{Representative Approach Analysis}
\label{sec:approach_analysis}

To understand the state of ML-based Android malware detection, a quantitative analysis of current work is indispensable.
While an ideal scenario is evaluating as many approaches as possible, conducting an exhaustive examination of each method is impractical due to the vast amount of existing literature.
Moreover, it is important to understand that, despite vast publications in this area achieving promising results, many of them share similar techniques (\textit{e.g.,} similar neural networks), and novel solutions are comparatively fewer.
Thus, our analysis focuses on methods that represent the broad spectrum and depth of advancements in the field.
That is, we aim to select representative approaches and utilize our framework to re-implement and evaluate them.
Specifically, in this section, we outline the selected approaches and provide a comparative analysis to highlight their experimental designs.
We then quantitatively evaluate these approaches in Section~\ref{sec:evaluation}, shedding further light on the current state of ML-based Android malware detection.

\subsection{Selection Criteria}
\label{sub:selection_criteria}

\bulletpoint{Cover various communities}
Android malware detection stands as an interdisciplinary domain, drawing contributions from diverse communities, including software engineering, security, and machine learning.
However, a notable separation is observed --- the solutions in one community often only compare with others from the same community --- which hinders potential collaborative advancements.
For instance, methods like~\cite{hou2017hindroid,ye2019out} from one community are often evaluated in isolation, complicating direct performance comparison.
To bridge the gap, we incorporate approaches from leading venues across a broad range of communities.

\bulletpoint{Explore an extensive spectrum of techniques in the detection pipeline}
As identified in Sec.~\ref{sec:systematic_investigation}, a wide range of technique combinations exists across different phases of the detection pipeline.
This study explores as many techniques as possible in each phase, including various mixes of APK features, feature representations, and ML models.
Such a comprehensive investigation enables us to gain a thorough understanding of the entire workflow of ML-based Android malware detection.

\bulletpoint{Reflect research progress}
In consideration of the evolving research landscape, where new methodologies continually emerge, we place emphasis on approaches that introduce cutting-edge techniques.
Particularly, we prioritize methods that either propose novel feature representations or employ innovative learning architectures, or achieve remarkable performance in the field.

\bulletpoint{Emphasize representative approaches over specific papers}
Many approaches share similar techniques, extracting patterns from analogous feature sets (\textit{e.g.,} program graphs) with similar ML models such as different variants of GNNs.
Analyzing these methods could lead to redundancy and provide limited insights.
Thus, we focus on distinct and representative strategies that offer more significant contributions.

\begin{table*}[t]
  \centering
  \caption{A summary of our selected approaches regarding APK Characterization, Feature Representation, and ML Models.
  \ymark \ indicates that the APK feature is utilized in feature engineering, while \nmark \ is the opposite.}
  \begin{adjustbox}{width=\linewidth, center}
  \begin{tabular}{@{}c|cccccccccc|c|cc@{}}
  \toprule
  \multirow{2}{*}{\textbf{\begin{tabular}[c]{@{}c@{}}\\Selected \\ Approach\end{tabular}}} & \multicolumn{10}{c|}{\textbf{APK Characterization}} & \multirow{2}{*}{\textbf{\begin{tabular}[c]{@{}c@{}} \\Feature \\ Representation\end{tabular}}} & \multirow{2}{*}{\textbf{\begin{tabular}[c]{@{}c@{}} \\ML \\ Models\end{tabular}}} \\ \cmidrule(lr){2-11}
    &\textbf{\begin{tabular}[c]{@{}c@{}}Hardware\\ Component\end{tabular}} & \textbf{\begin{tabular}[c]{@{}c@{}}Application\\ Component\end{tabular}} & \textbf{Intent} & \textbf{Permission} & \textbf{\begin{tabular}[c]{@{}c@{}}API\\ Call\end{tabular}} & \textbf{\begin{tabular}[c]{@{}c@{}}Byte\\ Code\end{tabular}} & \textbf{Opcode} & \textbf{\begin{tabular}[c]{@{}c@{}}Code\\ String\end{tabular}} & \textbf{\begin{tabular}[c]{@{}c@{}}Program\\ Graph\end{tabular}} & \textbf{\begin{tabular}[c]{@{}c@{}}Resource\\ Information\end{tabular}} &   \\ \midrule
  
  Drebin~\cite{arp2014drebin}&\ymark&\ymark&\ymark&\ymark&\ymark&\nmark&\nmark&\ymark&\nmark&\nmark& Categorical & SVM &  \\ \midrule
  MamaDroid~\cite{mariconti2016mamadroid} &\nmark&\nmark&\nmark&\nmark&\ymark&\nmark&\nmark&\nmark&\ymark&\nmark& Graph-based & RF &  \\ \midrule
  Mclaughlin et al.~\cite{mclaughlin2017deep} &\nmark&\nmark&\nmark&\nmark&\nmark&\nmark&\ymark&\nmark&\nmark&\nmark& Image-based & CNN  \\ \midrule
  HinDroid~\cite{hou2017hindroid} &\nmark&\nmark&\nmark&\nmark&\ymark&\nmark&\nmark&\nmark&\ymark&\nmark& Graph-based & SVM  \\ \midrule
  DeepRefiner~\cite{xu2018deeprefiner}  &\ymark&\ymark&\ymark&\ymark&\nmark&\ymark&\nmark&\nmark&\nmark&\ymark& Text-based & LSTM  \\ \midrule
  Kim et al.~\cite{kim2018multimodal} &\ymark&\ymark&\ymark&\ymark&\ymark&\nmark&\ymark&\ymark&\nmark&\nmark& Categorical & MLP   \\ \midrule
  MalScan~\cite{wu2019malscan} &\nmark&\nmark&\nmark&\nmark&\ymark&\nmark&\nmark&\nmark&\ymark&\nmark& Graph-based & KNN   \\ \midrule
  SDAC~\cite{xu2020sdac} &\nmark&\nmark&\nmark&\nmark&\ymark&\nmark&\nmark&\nmark&\ymark&\nmark& Categorical & SVM   \\ \midrule
  HomDroid~\cite{wu2021homdroid} &\nmark&\nmark&\nmark&\nmark&\ymark&\nmark&\nmark&\nmark&\ymark&\nmark& Categorical & KNN & \\ \midrule
  Xmal~\cite{wu2021android} &\nmark&\nmark&\nmark&\ymark&\ymark&\nmark&\nmark&\nmark&\nmark&\nmark& Categorical & MLP  \\ \midrule
  RAMDA~\cite{li2021robust} &\nmark&\nmark&\ymark&\ymark&\ymark&\nmark&\nmark&\nmark&\nmark&\nmark& Categorical & AE  \\ \midrule
  MSDroid~\cite{he2022msdroid} &\nmark&\nmark&\nmark&\ymark&\ymark&\nmark&\ymark&\ymark&\ymark&\nmark& Graph-based & GNN  \\ \bottomrule
  \end{tabular}
  \end{adjustbox}
  
  \centering
  \label{tbl:systematization}
  \begin{tablenotes}
  \footnotesize
  \item While multiple ML models may be utilized in individual approaches~\cite{mariconti2016mamadroid,wu2019malscan,wu2021homdroid}, we report the model that yields the best effectiveness (\textit{e.g.}, F1-score).
  \end{tablenotes}
  \label{tab:apk_characterization}
  \vspace{-0.5cm}
\end{table*}

\subsection{Selected Approaches}
\label{sub:selected_approaches}

In Section~\ref{sec:systematic_investigation}, we have presented recent advancements in ML-based Android malware detection by dissecting the detection pipeline into three phases: APK characterization, feature representation, and ML modeling.
Adhering to the selection criteria, we identify 12 representative approaches from hundreds of available solutions to analyze the current state of this field.
During the selection process, to foster collaboration among different fields, we prioritize the approaches published in leading venues from three communities, such as ASE, TOSEM from software engineering, NDSS, TIFS from security, and KDD, WWW from machine learning.
Specifically, the selected approaches include 3 from software engineering, 7 from security, and 2 from machine learning.
A detailed summary about sources and publication years of these approaches is provided in Table~\ref{tab:literature}.
We also ensure that the selected approaches cover almost all the APK characterization, feature representation, and ML models discussed in Section~\ref{sec:systematic_investigation}.
As Table~\ref{tab:apk_characterization} shows, they are carefully chosen to represent a broad range of techniques employed in the detection process, including 10 APK features, 4 feature representations, and 8 ML models.
Additionally, the selected approaches are distinguished either by their promising performance or by introducing novel techniques.
For instance, both MsDroid~\cite{he2022msdroid} and EFCG~\cite{cai2021learning} are recent works that introduce GNNs to detect malicious patterns in APKs.
We highlight MsDroid as it better represents apps' graph structures via aggregating more channel information and currently stands as the state-of-the-art in Android malware detection using GNNs.
Importantly, we also make a concerted effort to ensure that the selected solutions are not variants of existing ones.
For example, several methods~\cite{hou2017hindroid,ye2019out,fan2021heterogeneous} utilize heterogeneous graphs to model APKs, we select HinDroid~\cite{hou2017hindroid} because it introduces a novel feature representation that encodes diverse API call relationships while achieving performance comparable to other methods.
To better understand the selected representative approaches, we introduce them in the following section.

\bulletpoint{Drebin}
Drebin~\cite{arp2014drebin} first collects APK features, such as hardware components, permissions, intents, and API calls, using static analysis.
These features are then converted into a binary vector to signify their existence or not.
An SVM is then trained to detect Android malware.

\bulletpoint{MamaDroid}
MamaDroid~\cite{mariconti2016mamadroid} pioneers the use of Markov chains to model app behavior. 
It constructs a Markov chain over the sequence of abstracted API calls invoked by an app and computes the transition probabilities between these calls as features, which are then used by an RF classifier to detect Android malware.

\bulletpoint{Mclaughlin et al}
This technique~\cite{mclaughlin2017deep} presents the leading edge in image-based Android malware detection.
By transforming opcode sequences into images via a one-hot encoding, it leverages a CNN model to classify them as benign or malicious samples.

\bulletpoint{HinDroid}
Hou et al.~\cite{hou2017hindroid} introduce a novel approach by representing API calls as a structured heterogeneous information graph.
This approach accounts for the inter-relationships among API calls, such as their presence in the same code block.
It then captures apps' semantics with meta-path techniques~\cite{sun2011pathsim}.
A multi-kernel SVM is further applied to recognize malicious patterns.

\bulletpoint{DeepRefiner}
DeepRefiner~\cite{xu2018deeprefiner} designs a two-layer malware detection system.
Initially, it feeds features like hardware components, permissions, and resources into an MLP to detect most malware.
For ambiguous cases, it further interprets APK bytecodes as text sequences and employs an LSTM model to capture the method-level and application-level semantics behind app behaviors.

\bulletpoint{Kim et al}
Kim et al.~\cite{kim2018multimodal} use multi-modal learning to detect malware, aggregating various features, such as intent and API calls. 
Distinct MLPs are initially utilized to process individual features independently. 
Subsequently, a unified MLP integrates the outputs from the preceding models, offering a consolidated decision on malware identification.

\bulletpoint{MalScan}
This study~\cite{wu2019malscan} treats program graphs as social networks, where API calls are treated as nodes, and the relationships between them are presented as edges.
The system further evaluates the centrality of sensitive API calls in the graph to derive features and then feeds them into a KNN model to detect malware.

\bulletpoint{SDAC}
\jh{It attempts to cluster API calls based on their contextual information extracted from API call sequences~\cite{xu2020sdac}.}
These resulting clusters act as features to represent APKs. 
An SVM model is then used to capture malicious patterns.

\bulletpoint{HomDroid}
The method~\cite{wu2021homdroid} focuses on suspicious parts of malware by calculating the homophily in its program graph.
From the malicious subgraphs, it derives two key features: (1) the presence of sensitive API calls, and (2) the number of sensitive triads. 
\jh{These features are then fed into a KNN model to detect malware.}

\bulletpoint{Xmal}
Xmal~\cite{wu2021android} utilizes MLPs to distill information from extracted API calls and permissions for Android malware detection.
It further integrates an attention mechanism to highlight the most informative features.
This attention-based MLP not only achieves promising results but also offers an interpretation of the model.

\bulletpoint{RAMDA}
This detector~\cite{li2021robust} is the SOTA approach that employs Autoencoder to derive a resilient representation of APKs with features such as API calls and intents.
The representation is fed into an MLP to detect malware.

\bulletpoint{MSDroid}
MSDroid~\cite{he2022msdroid} is the cutting-edge in utilizing GNN to detect malware.
Initially, it breaks down the program graph into subgraphs rooted at sensitive API calls. 
Then, it leverages a GNN to capture essential semantics from the graph representations for malware detection.

\subsection{Comparative Study}
\label{sub:compare}

\begin{table*}[t]
  \centering
  \caption{A comparative study of our selected approaches based on their experimental setup, efficiency evaluation, robustness evaluation, artifact release, and toolchain. 
  --- denotes the absence of the statistics in the literature. 
  \ymark \ indicates that both feature encoding and ML modeling were evaluated for efficiency, \halfmark \ indicates that only ML modeling was evaluated, and \nmark \ indicates that the efficiency was not evaluated. 
  Malware Ratio refers to the proportion of malware samples in a testing set.}
  \vspace{-0.1cm}
  \begin{adjustbox}{width=\linewidth, center}
  
  \begin{tabular}{@{}c|rccc|c|ccc|c|c@{}}
  \toprule
  \multirow{2}{*}{\textbf{\begin{tabular}[c]{@{}c@{}}\\Selected \\ Approach\end{tabular}}} & \multicolumn{4}{c|}{\textbf{Experimental Setup}} & \multirow{2}{*}{\textbf{\begin{tabular}[c]{@{}c@{}}\\Efficiency\\Evaluation\end{tabular}}} & \multicolumn{3}{c|}{\textbf{Robustness Evaluation}} & \multirow{2}{*}{\textbf{\begin{tabular}[c]{@{}c@{}}\\Artifact \\ Release\end{tabular}}} & \multirow{2}{*}{\textbf{\begin{tabular}[c]{@{}c@{}}\\Tool \\ Chain\end{tabular}}} \\ \cmidrule(lr){2-5} \cmidrule(lr){7-9}
   & \textbf{\begin{tabular}[c]{@{}c@{}}Dataset \\ Size\end{tabular}} & \textbf{\begin{tabular}[c]{@{}c@{}}Time \\ Span\end{tabular}} & \textbf{\begin{tabular}[c]{@{}c@{}}Train:\\ Val:Test\end{tabular}} & \textbf{\begin{tabular}[c]{@{}c@{}}Malware\\Ratio\end{tabular}} &  & \textbf{Evolution} & \textbf{Obfuscation} & \textbf{\begin{tabular}[c]{@{}c@{}}Adversarial\\ Sample\end{tabular}} &  \\ \midrule
  Drebin~\cite{arp2014drebin} & 129,013 & 2010-2012 & 2:0:1 & 4\% & \ymark & \xmark & \xmark & \xmark & \cmark & Androguard \\ \midrule
  MamaDroid~\cite{mariconti2016mamadroid} & 43,940 & 2010-2016 & 9:0:1 & 50\% & \ymark & \cmark & \xmark & \xmark & \cmark &Soot\\ \midrule
  Mclaughlin et al.~\cite{mclaughlin2017deep} & 27,395 & --- & --- & 50\% & \halfmark & \xmark & \xmark & \xmark & \cmark & BackSmali \\ \midrule
  HinDroid~\cite{hou2017hindroid} & 2,334 & 2017-2017 & 4:0:1 & 60\% & \halfmark & \xmark & \xmark & \xmark & \xmark & APKTool\\ \midrule
  DeepRefiner~\cite{xu2018deeprefiner} & 110,440 & --- & \textbf{8:1:1} & 57\% & \ymark & \xmark & \cmark & \cmark & \xmark & APKTool\\ \midrule
  Kim et al.~\cite{kim2018multimodal} & 41,260 & --- & \textbf{3:1:1} & 50\% & \halfmark & \xmark & \cmark & \xmark & \xmark & APKTool\\ \midrule
  MalScan~\cite{wu2019malscan} & 30,715 & 2011-2018 & 9:0:1 & 50\% & \ymark & \cmark & \xmark & \cmark & \cmark & Androguard\\ \midrule
  SDAC~\cite{xu2020sdac} & 70,142 & 2011-2016 & 8:0:2 & 50\% & \ymark & \cmark & \cmark & \xmark & \xmark & FlowDroid\\ \midrule
  HomDroid~\cite{wu2021homdroid} & 8,198 & --- & 9:0:1 & 40\% & \ymark & \xmark & \xmark & \xmark & \xmark & Androguard\\ \midrule
  Xmal~\cite{wu2021android} & 35,690 & --- & 7:0:3 & 43\% & \nmark & \xmark & \xmark & \xmark & \cmark & Androguard\\ \midrule
  RAMDA~\cite{li2021robust} & 58,483 & --- & 19:0:1 & 50\% & \nmark & \xmark & \xmark & \cmark & \cmark & APKTool\\ \midrule
  MSDroid~\cite{he2022msdroid} & 81,790 & 2010-2015 & 4:0:1 & 37\% & \nmark & \cmark & \cmark & \xmark & \cmark & Androguard\\ \bottomrule
  \end{tabular}%
  
  \end{adjustbox}
  \label{tab:comparative_study}
  \vspace{-0.4cm}
\end{table*}

Analyzing these approaches from various angles provides insightful lens for understanding the current advancements and challenges in ML-based Android malware detection.
In this section, we conduct a comparative review of the selected approaches, focusing on the following three critical dimensions.
(a) \textit{Effectiveness} refers to the ability of an approach to accurately identify malware under various circumstances, such as different dataset sizes and goodware-to-malware ratios;
(b) \textit{Robustness} assesses the methods' resilience, especially in response to challenges like malware evolution;
(c) \textit{Efficiency} reflects the computational overhead incurred during APK processing and ML modeling.
Building on this, we further emphasize the importance of employing a unified framework to evaluate the effectiveness, robustness, and efficiency of ML-based Android malware detection techniques.

\bulletpoint{Effectiveness}
Effectiveness is the most important criterion for any detection technique, and all the selected approaches evaluate this aspect.
However, the experimental setup varies dramatically across these approaches, making direct comparisons challenging.
As indicated in Table~\ref{tab:comparative_study}, the experimental settings --- dataset size, time span, dataset partition, and malware ratio --- demonstrate large variations across studies.
\jh{It is well-established that a positive correlation exists between the training data size and ML model performance.}
The training data size in these approaches ranges from 2,334~\cite{hou2017hindroid} to 129,013~\cite{arp2014drebin}, making it difficult to compare their effectiveness.
The datasets' temporal span further complicates the evaluation, as Android malware evolves over time and the features extracted from APKs change accordingly.
Another point is the absence of a validation set~\cite{he2022msdroid,li2021robust,wu2021android}.
This oversight is alarming, raising concerns about potential over-fitting and over-optimistic performance indicators.
In the wild, the ratio of malware to benign apps is notably imbalanced, with malware accounting for around 10\% of cases~\cite{pendlebury2019tesseract}.
However, this ratio in testing datasets significantly varies across different studies (\textit{e.g.,} from 4\%~\cite{arp2014drebin} to 60\%~\cite{hou2017hindroid}), which makes it difficult to reveal the true effectiveness of these approaches.

\bulletpoint{Robustness}
Robustness stands as a pivotal criterion for any methodology. 
Within Android malware detection, this robustness typically encompasses an approach's capacity to counteract malware evolution, obfuscation strategies, and adversarial attacks~\cite{pendlebury2019tesseract}.
Specifically, detectors routinely operate in hostile and dynamic contexts~\cite{barbero2022transcending}, where malware constantly evolves to evade detection.
Also, obfuscation techniques have been widely adopted by attackers to conceal their malicious operations~\cite{aonzo2020obfuscapk}.
Additionally, the inherent susceptibility of ML models to adversarial attacks~\cite{li2023black,athalye2018obfuscated} complicates the detection process.
We observe that these selected approaches miss one or more of the challenges, as shown in Table~\ref{tab:comparative_study}.
This omission complicates the assessment of their true effectiveness in real-world deployments.

\bulletpoint{Efficiency}
To measure a new malware detector, the importance of efficiency stands parallel to effectiveness.
As Android apps grow in size and complexity, the time and computational resources required for APK processing and ML modeling could substantially rise.
However, as Table~\ref{tab:comparative_study} shows, not every approach evaluates the efficiency of these two parts.
Additionally, understanding how efficiency shifts when dealing with APKs at various times is vital to ensure detectors' sustainability and long-term utility; unfortunately, this is not considered by any of the selected approaches.

\bulletpoint{Artifact Release and Toolchain}
Table~\ref{tab:comparative_study} also reports whether the selected approaches make their artifacts publicly available and details the specific toolchains they employ. 
We observe that nearly half of these approaches do not release their artifacts, which poses significant challenges for reproducibility and subsequent research. 
Furthermore, the toolchains they use are diverse, including Androguard~\cite{androguard}, APKTool~\cite{apkt}, and BackSmali~\cite{backsmali}.
We know that different toolchains may yield varying analysis results for the same APK due to differences in their static analysis capabilities~\cite{arp2014drebin,mariconti2016mamadroid}.
As a result, it is difficult to fairly compare the effectiveness and efficiency of these approaches when they rely on heterogeneous toolchains.
To further investigate this issue, we conduct an experimental analysis in Section~\ref{sec:toolchain_analysis}.

Combining the aforementioned analyses, there is a pressing need for a general-purpose framework that standardizes the entire development pipeline, from feature extraction and model construction to training and deployment.
Such a framework should also support diverse and configurable evaluation scenarios for ML-based Android malware detectors, enabling fair comparison across approaches, improving reproducibility, and facilitating large-scale empirical studies.

\section{Quantitative Analysis}
\label{sec:evaluation}

One of our key contributions is a comprehensive quantitative analysis of the selected approaches.
This analysis aims to assess and disentangle the effects of various experimental settings --- such as data size, goodware-to-malware ratios, and the presence of adversarial attacks --- on the performance of these representative detectors.
By systematically varying these conditions, we can observe how current ML-based Android malware detection methods behave under different scenarios and identify key factors that contribute to an effective detector.
Specifically, we re-implement the 12 representative approaches outlined in Section~\ref{sec:approach_analysis} within our general-purpose framework, \codename\ (implementation details are provided in Appendix~\ref{sub:approach_overview}).
We then evaluate their effectiveness (Section.~\ref{effectiveness}), robustness (Section.~\ref{robustness}), and efficiency (Section.~\ref{efficiency}) on a large-scale, long-duration dataset (Section.~\ref{sec:dataset}) tailored for this study.
Beyond these experiments, we also investigate additional aspects, such as the influence of reverse-engineering toolchains and model selection given specific feature sets, to derive deeper insights into the current state of Android malware detection.

\subsection{Dataset}
\label{sec:dataset}

\bulletpoint{Dataset Construction Principles}
To conduct an evaluation that is representative of real-world scenarios and enables a comprehensive, multi-dimensional assessment, it is essential that the dataset satisfies the following criteria~\cite{pendlebury2019tesseract}.
\begin{itemize}[leftmargin=14pt]
    \item \textit{Market Diversity}: Android apps should be collected from multiple app stores to ensure comprehensive representation of the ecosystem. We consider various sources, including Google Play, VirusShare, and malware repositories.
    \item \textit{Exclusion of Grayware}: Grayware should be excluded to prevent skewing the results, as its uncertain nature can complicate the classification of apps. To filter out grayware, we use the number of positive antivirus alerts (denoted as $p$) derived from VirusTotal~\cite{virustotal} as a criterion. Following previous studies~\cite{pendlebury2019tesseract,miller2016reviewer}, apps with $p\geq 4$ are categorized as malicious, while those with $p=0$ are considered benign.
    \item \textit{Temporal Distribution}: Malware evolution is a critical factor affecting detection performance; thus, the dataset should span a wide range of years to capture this dynamic.
    \item \textit{Realistic Malware Ratio}: The dataset should reflect the actual goodware-to-malware ratio observed in the wild to ensure that the evaluation is grounded in real-world conditions. Specifically, we follow previous studies~\cite{pendlebury2019tesseract,chen2023continuous} and set the malware ratio to 10\%, to establish a realistic evaluation environment.
    \item \textit{Deduplication}: To avoid bias in the evaluation, duplicate apps should be removed from the dataset. We first perform hash-based deduplication to eliminate exact copies of the same APK. We then filter out apps that share the same package name and version code, as they are likely to be identical or highly similar. Finally, we inspect the extracted features, such as the number of permissions and API calls, to identify potential duplicates that may have been repackaged or slightly modified.
\end{itemize}

\begin{table}[t]
    \renewcommand{\arraystretch}{0.9}
    \centering
    \caption{Evaluation Dataset Statistics of the Primary Dataset.
    The unit used for measuring APK size is megabytes (MB).}
    \begin{adjustbox}{width=0.64\linewidth, center}
    \begin{tabular}{c|ccccc}
    \hline
    \textbf{Year}  & \textbf{Malicious} & \textbf{Benign} & \textbf{M+B} & \textbf{M/(M+B)} & \textbf{Avg.Size} \\ \hline
    \textbf{2011}  & 2,085   & 17,878   & 19,963& 10.4\%& 2.26   \\
    \textbf{2012}  & 2,137   & 18,687   & 20,824& 10.3\%& 3.58   \\
    \textbf{2013}  & 2,182   & 18,631   & 20,813& 10.5\%& 5.21   \\
    \textbf{2014}  & 2,346   & 20,142   & 22,488& 10.4\%& 6.91   \\
    \textbf{2015}  & 2,369   & 20,643   & 23,012& 10.3\%& 9.52   \\
    \textbf{2016}  & 2,390   & 21,292   & 23,682& 10.1\%& 12.26  \\
    \textbf{2017}  & 2,389   & 21,006   & 23,395& 10.2\%& 16.24  \\
    \textbf{2018}  & 2,326   & 20,099   & 22,425& 10.4\%& 16.62  \\
    \textbf{2019}  & 2,345   & 20,260   & 22,605& 10.4\%& 17.27  \\
    \textbf{2020}  & 2,301   & 19,802   & 22,103& 10.4\%& 16.65  \\ \hline
    \textbf{Total} & 22,870  & 198,440  & 221,310      & 10.3\%& 10.86  \\ \hline
    \end{tabular}
    \end{adjustbox}
    \label{tab:dataset_stat}
    \vspace{-0.3cm}
\end{table}

Following these principles, we construct two datasets: a primary dataset (Section~\ref{primary_dataset}) used for the main evaluation, and an additional dataset (Section~\ref{additional_dataset}) employed for further validation and complementary experiments.

\begin{table}[t]
    \centering
    \caption{Data distribution of four sub-datasets derived from primary dataset. M/N indicates M benign and N malicious apps.}
    \begin{adjustbox}{width=0.51\linewidth, center}
 \begin{tabular}{r|r|r|c}
     \hline
     \multicolumn{1}{c|}{\textbf{Training}} & \multicolumn{1}{c|}{\textbf{Validation}} & \multicolumn{1}{c|}{\textbf{Testing}} & \textbf{Alias} \\ \hline
     \multirow{2}{*}{14,400/1,600}    & 1,800/200    & 1,800/200 & \ding{182}  \\ \cline{2-4} 
   & 1,000/1,000  & 1,000/1,000      & \ding{183}  \\ \hline
     \multirow{2}{*}{8,000/8,000}     & 1,800/200    & 1,800/200 & \ding{184} \\ \cline{2-4} 
   & 1,000/1,000  & 1,000/1,000      &  \ding{185}\\ \hline
     \end{tabular}
    \end{adjustbox}
    \label{tab:effectiveness_dataset}
    \vspace{-0.3cm}
\end{table}

\subsubsection{\textbf{Primary Dataset}}
\label{primary_dataset}

To build the main evaluation dataset, we take AndroZoo~\cite{allix2016androzoo} as our app source, which is a growing collection of apps from different platforms, like Google Play, PlayDrone, and AppChina.
As for the temporal distribution, we set the time span as ten years (\textit{i.e.,} from 2011 to 2020) for a comprehensive evaluation, following previous studies~\cite{chen2023continuous,gao2024comprehensive}.
To mimic actual conditions, we select apps based on a monthly time window, ensuring that around 2,000 apps are collected each month, with a goodware-to-malware ratio aligning with the estimated $9:1$ in the wild~\cite{pendlebury2019tesseract}.
Additionally, to ensure the quality of the dataset, we exclude apps that cannot be parsed by AndroGuard~\cite{androguard} because we cannot extract the required features for malware detection.
Finally, we obtain a dataset of 221,310 Android apps, i.e., 22,870 malicious and 198,440 benign apps, as summarized in Table~\ref{tab:dataset_stat}.
We employ this primary dataset for the main evaluation of the selected approaches in effectiveness, robustness, and efficiency.

\bulletpoint{Settings}
To further support specific evaluation scenarios, we construct four sub-datasets with different configurations derived from the primary dataset.
For each sub-dataset, apps are randomly sampled to form training, validation, and testing sets, and their distributions are summarized in Table~\ref{tab:effectiveness_dataset}.
Importantly, the apps in these sub-datasets are sampled from different years to ensure comprehensive temporal representation.
For instance, for \ding{182}, each year contributes 1,440 benign and 160 malicious apps for training, 180 benign and 20 malicious apps for validation, and the same number for testing. 
Duplicate apps are also avoided across these sets.
We use these sub-datasets to investigate the effectiveness of the selected representative approaches under different settings, such as goodware-to-malware ratio and APK features.

\begin{figure}[t]
    \centering
    \includegraphics[width=0.64\linewidth]{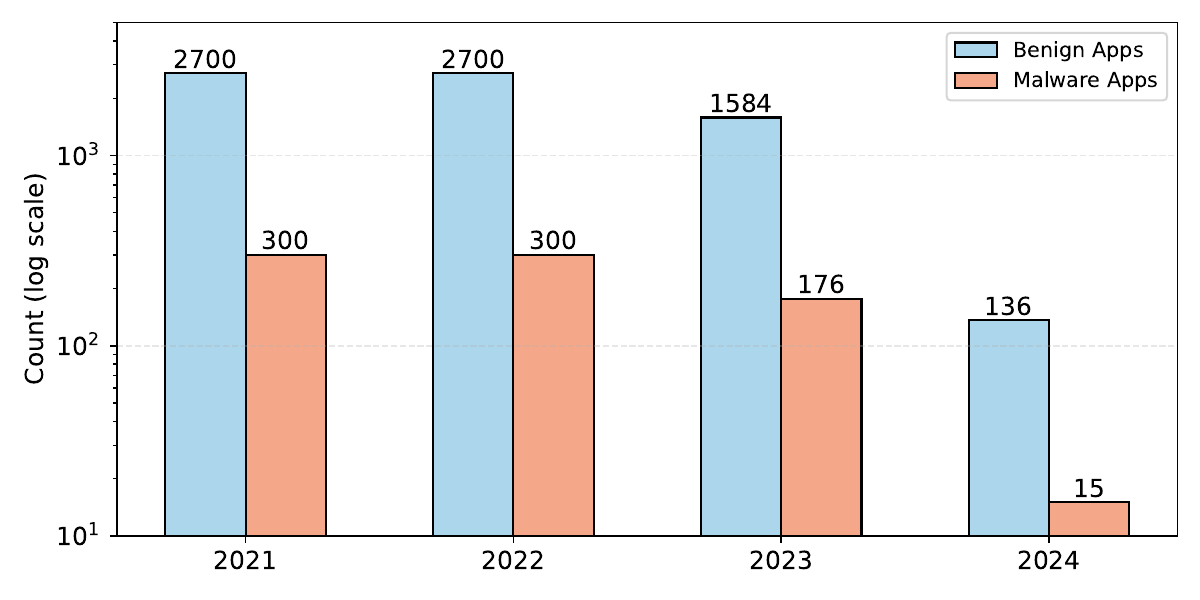}
    \vspace{-0.2cm}
    \caption{Data statistics of apps collected from 2021 to 2024. The y-axis uses a logarithmic scale.}
    \label{fig:new_data}
    \vspace{-0.3cm}
\end{figure}

\subsubsection{\textbf{Additional Dataset}}
\label{additional_dataset}

As a complementary dataset, we collect Android apps from different sources, \textit{i.e.,} AndroZoo~\cite{allix2016androzoo} and VirusTotal~\cite{virustotal}, spanning the years 2021 to 2024, following the same principles outlined above.
During this process, we observe that the number of malware samples in these years is relatively small, making it challenging to support a thorough evaluation~\cite{arp2022and}.
To enrich the dataset, we therefore additionally gather malware samples from VirusShare~\cite{vs}, a well-known repository that aggregates a large collection of malware from various sources.
We maintain a realistic goodware-to-malware ratio of 9:1 in this additional dataset.
Figure~\ref{fig:new_data} summarizes its statistics, which in total contains 7,911 apps, including 791 malicious and 7,120 benign samples.
We note that the number of malicious samples in this additional dataset is still relatively small compared to the primary dataset.
Nonetheless, it is sufficient to support our complementary experiments, which primarily aim to validate the findings from the main evaluation, particularly regarding the effectiveness (\textit{e.g.,} whether detection performance changes significantly) and efficiency (\textit{e.g.,} memory usage) of the selected approaches.

\subsection{Effectiveness}
\label{effectiveness}
In this section, we examine how the detection effectiveness of the selected approaches varies under different experimental settings, with the goal of identifying the key factors that contribute to an effective malware detector.
Specifically, we focus on the following aspects: goodware-to-malware ratio, training set size, APK feature choices, and model selection.
These factors are pivotal because they influence the quality of the training data and the model's ability to extract relevant information, both of which are critical to the performance of ML-based detectors.
All experiments in this section are conducted on the primary dataset (Section~\ref{sec:dataset}).

\begin{table}[t]
    \centering
    \caption{The effectiveness (F1 and Accuracy) of the approaches across varied goodware-to-malware ratios in training and testing sets.}
    \begin{adjustbox}{width=0.78\linewidth, center}
 \begin{tabular}{@{}c|cccc|cccc@{}}
     \toprule
     \multirow{3}{*}{\textbf{\makecell[c]{Selected\\ Approach}}} & \multicolumn{4}{c|}{\textbf{B:M=9:1 in Tr (\ding{182} - \ding{183})}}& \multicolumn{4}{c}{\textbf{B:M=1:1 in Tr (\ding{184} - \ding{185})}}\\ \cmidrule(l){2-9} 
     & \multicolumn{2}{c|}{\textbf{B:M=9:1 in Ts}}& \multicolumn{2}{c|}{\textbf{B:M=1:1 in Ts}}& \multicolumn{2}{c|}{\textbf{B:M=9:1 in Ts}}& \multicolumn{2}{c}{\textbf{B:M=1:1 in Ts}} \\ \cmidrule(l){2-9} 
     & \multicolumn{1}{c|}{\textbf{F1}} & \multicolumn{1}{c|}{\textbf{Acc.}} & \multicolumn{1}{c|}{\textbf{F1}} & \textbf{Acc.} & \multicolumn{1}{c|}{\textbf{F1}} & \multicolumn{1}{c|}{\textbf{Acc.}} & \multicolumn{1}{c|}{\textbf{F1}} & \textbf{Acc.} \\ \midrule
     Drebin & \multicolumn{1}{c|}{0.722}& \multicolumn{1}{c|}{\textbf{0.948}} & \multicolumn{1}{c|}{0.791}& 0.823 & \multicolumn{1}{c|}{0.650}& \multicolumn{1}{c|}{0.901} & \multicolumn{1}{c|}{0.918}& 0.917 \\ \midrule
     MamaDroid& \multicolumn{1}{c|}{0.661}& \multicolumn{1}{c|}{0.945} & \multicolumn{1}{c|}{0.693}& 0.763 & \multicolumn{1}{c|}{0.650}& \multicolumn{1}{c|}{0.902} & \multicolumn{1}{c|}{0.895}& 0.897 \\ \midrule
     Mclaughlin et al.& \multicolumn{1}{c|}{0.714}& \multicolumn{1}{c|}{0.946} & \multicolumn{1}{c|}{0.799}& 0.828 & \multicolumn{1}{c|}{0.682}& \multicolumn{1}{c|}{\textbf{0.924}} & \multicolumn{1}{c|}{0.916}& 0.916 \\ \midrule
     HinDroid & \multicolumn{1}{c|}{\textbf{0.731}}& \multicolumn{1}{c|}{0.943} & \multicolumn{1}{c|}{0.819}& 0.842 & \multicolumn{1}{c|}{\textbf{0.701}}& \multicolumn{1}{c|}{0.924} & \multicolumn{1}{c|}{\textbf{0.925}}& \textbf{0.925} \\ \midrule
     DeepRefiner & \multicolumn{1}{c|}{0.657}& \multicolumn{1}{c|}{0.932} & \multicolumn{1}{c|}{0.776}& 0.809 & \multicolumn{1}{c|}{0.667}& \multicolumn{1}{c|}{0.918} & \multicolumn{1}{c|}{0.881}& 0.881 \\ \midrule
     Kim et al.& \multicolumn{1}{c|}{\textbf{0.782}}& \multicolumn{1}{c|}{\textbf{0.952}} & \multicolumn{1}{c|}{\textbf{0.907}}& \textbf{0.912} & \multicolumn{1}{c|}{\textbf{0.753}}& \multicolumn{1}{c|}{0.\textbf{941}} & \multicolumn{1}{c|}{\textbf{0.938}}& \textbf{0.937} \\ \midrule
     MalScan& \multicolumn{1}{c|}{0.684}& \multicolumn{1}{c|}{0.939} & \multicolumn{1}{c|}{0.793}& 0.823 & \multicolumn{1}{c|}{0.587}& \multicolumn{1}{c|}{0.877} & \multicolumn{1}{c|}{0.880}& 0.880 \\ \midrule
     SDAC& \multicolumn{1}{c|}{0.522}& \multicolumn{1}{c|}{0.916} & \multicolumn{1}{c|}{0.627}& 0.720 & \multicolumn{1}{c|}{0.524}& \multicolumn{1}{c|}{0.850} & \multicolumn{1}{c|}{0.845}& 0.844 \\ \midrule
     HomDroid & \multicolumn{1}{c|}{\textbf{0.734}}& \multicolumn{1}{c|}{\textbf{0.949}} & \multicolumn{1}{c|}{0.816}& 0.841 & \multicolumn{1}{c|}{\textbf{0.701}}& \multicolumn{1}{c|}{\textbf{0.925}} & \multicolumn{1}{c|}{0.912}& 0.914 \\ \midrule
     Xmal& \multicolumn{1}{c|}{0.698}& \multicolumn{1}{c|}{0.942} & \multicolumn{1}{c|}{0.826}& \textbf{0.847} & \multicolumn{1}{c|}{0.674}& \multicolumn{1}{c|}{0.916} & \multicolumn{1}{c|}{\textbf{0.923}}& \textbf{0.924} \\ \midrule
     RAMDA& \multicolumn{1}{c|}{0.636}& \multicolumn{1}{c|}{0.905} & \multicolumn{1}{c|}{\textbf{0.841}}& \textbf{0.852} & \multicolumn{1}{c|}{0.510}& \multicolumn{1}{c|}{0.829} & \multicolumn{1}{c|}{0.871}& 0.865 \\ \midrule
     MSDroid& \multicolumn{1}{c|}{0.648}& \multicolumn{1}{c|}{0.919} & \multicolumn{1}{c|}{\textbf{0.834}}&0.828 & \multicolumn{1}{c|}{0.522}& \multicolumn{1}{c|}{0.853} & \multicolumn{1}{c|}{0.867}& 0.858 \\ \bottomrule
     \end{tabular}
    \end{adjustbox}
    \label{tab:res_ratio}
    \begin{tablenotes}[flushleft, labelsep=0]
 \footnotesize
 \item \textit{Tr:} training set, \textit{Ts:} testing set. \ding{182} - \ding{185} correspond to the four scenarios in Table~\ref{tab:effectiveness_dataset}. The top 3 results for each scenario are highlighted in bold.
 \end{tablenotes}
    \vspace{-0.5cm}
\end{table}

\bulletpoint{Goodware-to-malware ratio}
The effectiveness of ML-based Android malware detectors is heavily influenced by malware distributions in the training and testing sets.
Here, we consider two ratios, \textit{i.e.}, 10\%, and 50\%, in both training and testing sets.
These choices are inspired by the estimated 10\% in the wild~\cite{pendlebury2019tesseract,chen2023continuous} and the 50\% widely used in previous studies~\cite{wu2019malscan,li2021robust,mariconti2016mamadroid}.
The \ding{182} - \ding{185} in Table~\ref{tab:effectiveness_dataset} present these four scenarios, where \ding{182} and \ding{183} have a 10\% malware ratio in training sets, while \ding{184} and \ding{185} have a 50\% malware ratio.

Table~\ref{tab:res_ratio} summarizes the results in terms of F1-score and Accuracy.
The results are arranged from left to right, corresponding to scenarios \ding{182} through \ding{185}.
We observe that all approaches achieve promising performance in all metrics under scenario \ding{185}, which is characterized by a \texttt{1:1} goodware-to-malware ratio in both the training and testing sets.
These findings are consistent with the results reported in the original papers, confirming the validity of our re-implementations. 
This demonstrates that these approaches can effectively detect Android malware under ideal conditions, where goodware and malware are balanced.
However, when the malware proportion is reduced to 10\% (\ding{182}), there is a pronounced decline in F1-score across all methods.
It is evident that many existing approaches experience performance degradation when evaluated under realistic malware ratios.
This is primarily attributed to the scarcity of malware samples in the training data, which constrains the model's ability to learn discriminative malware semantics.
Moreover, when the malware ratio in the test set is increased from 10\% to 50\% (\ding{183} and \ding{185}), we observe a consistent improvement in F1-scores across all methods.
This trend is expected, as a higher prevalence of malware samples in the test set increases the likelihood that models will encounter patterns similar to those seen during training, enhancing their predictive performance.
To more clearly reveal the performance differences between deep learning (DL)-based and traditional machine learning (TML)-based approaches, we conduct a coarse-grained, averaged analysis of the results, as averaging can partially mitigate the influence of the differing feature combinations employed by both types of approaches.
We observe an interesting phenomenon: DL-based approaches appear to outperform TML-based methods when the testing malware ratio is 50\% (\ding{183} and \ding{185}) --- almost all the top three results are achieved by DL-based methods.
While in more realistic scenarios (\ding{182} and \ding{184}), DL-based approaches do not exhibit a pronounced advantage over TML-based methods.
One possible explanation is that DL-based methods are better equipped to capture the complex semantics underlying app behaviors, but require more malware samples to effectively distill such semantics, which tend to be more prominent in datasets with a higher malware ratio.

\begin{table}[t]
    \centering
    \renewcommand{\arraystretch}{0.9}
    \caption{The effectiveness (precision and recall) of the approaches across varied goodware-to-malware ratios.}
    \vspace{-0.1cm}
    \begin{adjustbox}{width=0.84\linewidth, center}
    \begin{tabular}{@{}c|cccc|cccc@{}}
     \toprule
    \multirow{3}{*}{\textbf{\makecell[c]{Selected\\ Approach}}}
     & \multicolumn{4}{c|}{\textbf{B:M=9:1 in Tr (\ding{182} - \ding{183})}}
     & \multicolumn{4}{c}{\textbf{B:M=1:1 in Tr (\ding{184} - \ding{185})}}\\ 
     \cmidrule(l){2-9} 
     & \multicolumn{2}{c|}{\textbf{B:M=9:1 in Ts}}
     & \multicolumn{2}{c|}{\textbf{B:M=1:1 in Ts}}
     & \multicolumn{2}{c|}{\textbf{B:M=9:1 in Ts}}
     & \multicolumn{2}{c}{\textbf{B:M=1:1 in Ts}} \\ 
     \cmidrule(l){2-9} 
     & \multicolumn{1}{c|}{\textbf{Prec.}} 
     & \multicolumn{1}{c|}{\textbf{Rec.}} 
     & \multicolumn{1}{c|}{\textbf{Prec.}} 
     & \textbf{Rec.} 
     & \multicolumn{1}{c|}{\textbf{Prec.}} 
     & \multicolumn{1}{c|}{\textbf{Rec.}} 
     & \multicolumn{1}{c|}{\textbf{Prec.}} 
     & \textbf{Rec.} \\ 
     \midrule

     Drebin          & \multicolumn{1}{c|}{0.776} & \multicolumn{1}{c|}{0.675} & \multicolumn{1}{c|}{0.964} & 0.671 & \multicolumn{1}{c|}{0.501} & \multicolumn{1}{c|}{0.925} & \multicolumn{1}{c|}{0.906} & 0.930 \\ \midrule
     MamaDroid       & \multicolumn{1}{c|}{0.863} & \multicolumn{1}{c|}{0.535} & \multicolumn{1}{c|}{0.984} & 0.535 & \multicolumn{1}{c|}{0.506} & \multicolumn{1}{c|}{0.910} & \multicolumn{1}{c|}{0.905} & 0.886 \\ \midrule
     Mclaughlin et al.& \multicolumn{1}{c|}{0.758} & \multicolumn{1}{c|}{0.675} & \multicolumn{1}{c|}{0.958} & 0.685 & \multicolumn{1}{c|}{0.584} & \multicolumn{1}{c|}{0.820} & \multicolumn{1}{c|}{0.908} & 0.925 \\ \midrule
     HinDroid        & \multicolumn{1}{c|}{0.722} & \multicolumn{1}{c|}{0.740} & \multicolumn{1}{c|}{0.956} & 0.717 & \multicolumn{1}{c|}{0.578} & \multicolumn{1}{c|}{0.890} & \multicolumn{1}{c|}{0.914} & 0.937 \\ \midrule
     DeepRefiner     & \multicolumn{1}{c|}{0.663} & \multicolumn{1}{c|}{0.650} & \multicolumn{1}{c|}{0.944} & 0.659 & \multicolumn{1}{c|}{0.562} & \multicolumn{1}{c|}{0.820} & \multicolumn{1}{c|}{0.883} & 0.878 \\ \midrule
     Kim et al.      & \multicolumn{1}{c|}{0.717} & \multicolumn{1}{c|}{0.860} & \multicolumn{1}{c|}{0.960} & 0.859 & \multicolumn{1}{c|}{0.644} & \multicolumn{1}{c|}{0.905} & \multicolumn{1}{c|}{0.922} & 0.954 \\ \midrule
     MalScan         & \multicolumn{1}{c|}{0.704} & \multicolumn{1}{c|}{0.665} & \multicolumn{1}{c|}{0.954} & 0.678 & \multicolumn{1}{c|}{0.442} & \multicolumn{1}{c|}{0.875} & \multicolumn{1}{c|}{0.878} & 0.881 \\ \midrule
     SDAC            & \multicolumn{1}{c|}{0.632} & \multicolumn{1}{c|}{0.480} & \multicolumn{1}{c|}{0.908} & 0.511 & \multicolumn{1}{c|}{0.379} & \multicolumn{1}{c|}{0.850} & \multicolumn{1}{c|}{0.835} & 0.838 \\ \midrule
     HomDroid        & \multicolumn{1}{c|}{0.755} & \multicolumn{1}{c|}{0.714} & \multicolumn{1}{c|}{0.963} & 0.708 & \multicolumn{1}{c|}{0.583} & \multicolumn{1}{c|}{0.879} & \multicolumn{1}{c|}{0.926} & 0.898 \\ \midrule
     Xmal            & \multicolumn{1}{c|}{0.728} & \multicolumn{1}{c|}{0.670} & \multicolumn{1}{c|}{0.953} & 0.729 & \multicolumn{1}{c|}{0.549} & \multicolumn{1}{c|}{0.875} & \multicolumn{1}{c|}{0.927} & 0.920 \\ \midrule
     RAMDA           & \multicolumn{1}{c|}{0.516} & \multicolumn{1}{c|}{0.830} & \multicolumn{1}{c|}{0.910} & 0.781 & \multicolumn{1}{c|}{0.357} & \multicolumn{1}{c|}{0.890} & \multicolumn{1}{c|}{0.829} & 0.918 \\ \midrule
     MSDroid         & \multicolumn{1}{c|}{0.563} & \multicolumn{1}{c|}{0.760} & \multicolumn{1}{c|}{0.805} & 0.866 & \multicolumn{1}{c|}{0.387} & \multicolumn{1}{c|}{0.800} & \multicolumn{1}{c|}{0.820} & 0.919 \\

     \bottomrule
     \end{tabular}
    \end{adjustbox}
    \label{tab:res_ratio_prec_rec}
    \begin{tablenotes}[flushleft, labelsep=0]
    \footnotesize
    \item \textit{Tr:} training set, \textit{Ts:} testing set. \ding{182} - \ding{185} correspond to the four scenarios in Table~\ref{tab:effectiveness_dataset}. The top 3 results for each scenario are highlighted in bold.
    \end{tablenotes}
    \vspace{-0.4cm}
\end{table}

Table~\ref{tab:res_ratio_prec_rec} provides Precision and Recall values for these scenarios.
When the training set ratio is fixed, we observe that the malware detection performance (in terms of Recall) does not change significantly as the testing set ratio varies, as shown by pairs \ding{182}-\ding{183} and \ding{184}-\ding{185} in Table~\ref{tab:res_ratio_prec_rec}. 
In contrast, when the testing set ratio is fixed, we find that the Recall improves as the training set ratio increases from 10\% to 50\%, as illustrated by pairs \ding{182}-\ding{184} and \ding{183}-\ding{185} in Table~\ref{tab:res_ratio_prec_rec}.
This indicates that having a higher proportion of malware samples in the training set enhances the model's ability to learn effective malicious patterns, thereby improving detection accuracy.

\begin{figure}[t]
    \centering
    \includegraphics[width=0.74\linewidth]{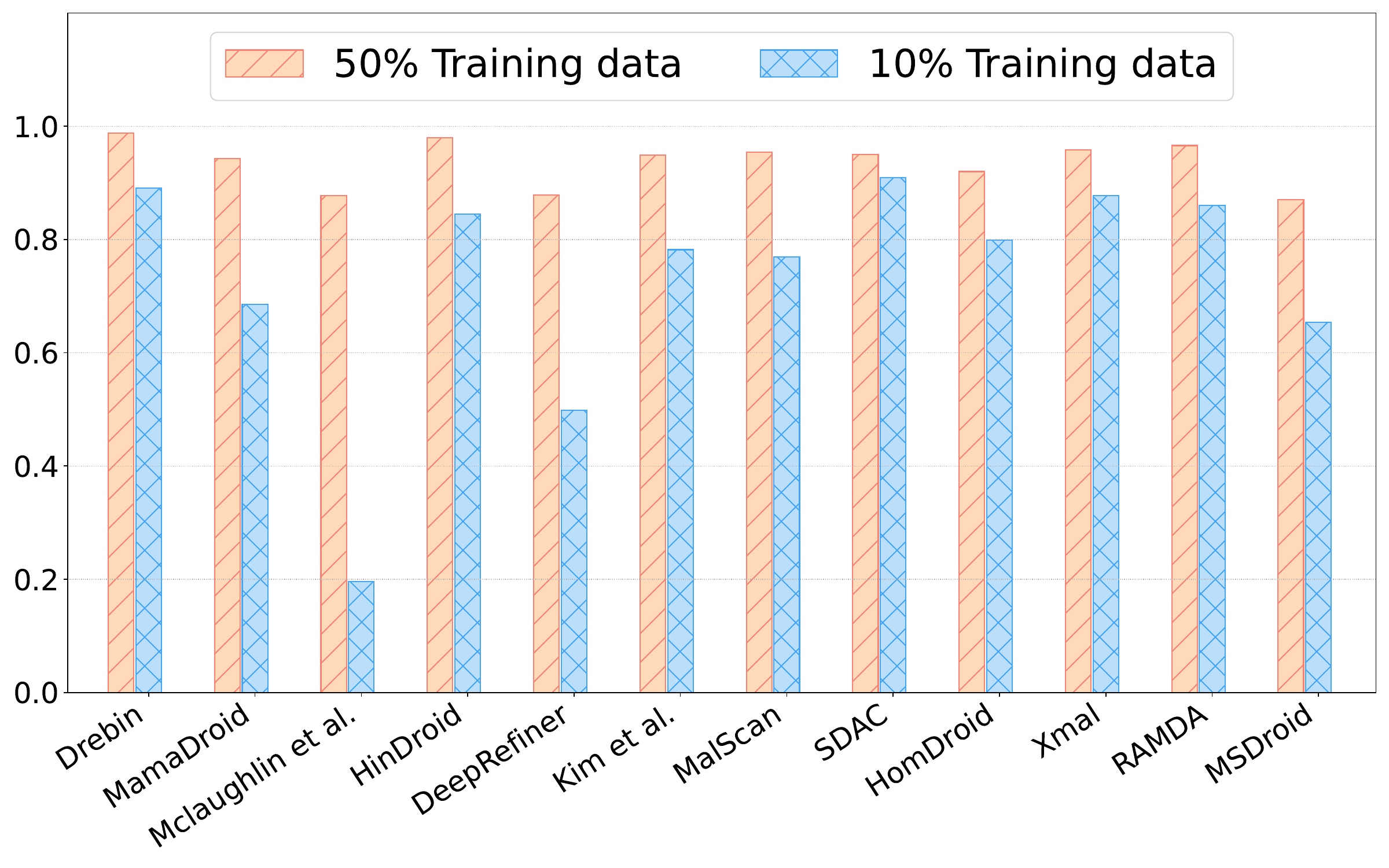}
    \vspace{-0.2cm}
    \caption{Effectiveness of the selected approaches using different sizes of training data.}
    \label{fig:training_size}
    \vspace{-0.5cm}
\end{figure}

\bulletpoint{Training set size}
The size of the training set is a critical determinant of the effectiveness of ML classifiers~\cite{sun2017revisiting}, as larger, high-quality datasets typically provide richer semantic information about malware, which is essential for learning effective classifiers.
However, obtaining such large, high-quality training sets is often infeasible due to the substantial costs associated with data collection and labeling.
In this experiment, we experiment with different training set sizes to simulate varying levels of semantic richness (\textit{i.e.,} larger datasets are assumed to encompass more diverse malware semantics) and analyze their impact on classifier performance.
Specifically, we consider two training set sizes, 50\% and 10\% of the original training set size, while maintaining the malware ratio at 10\%.

Figure~\ref{fig:training_size} shows how these detectors' performance changes when the training set size or semantic richness is reduced.
We normalize the F1-scores of these methods based on their performance obtained with the entire original dataset.
The detailed results are provided in the Appendix~\ref{sec:additional_results}.
The figure clearly shows that reducing the training set size leads to performance degradation across all approaches, underscoring more data enhances capturing malicious patterns.
Interestingly, Mclaughlin et al. and DeepRefiner display a greater sensitivity to training set size compared to others.
One explanation is that, unlike other methods that heuristically select features from apps, these two approaches take the original apps' bytecode as input and automatically extract semantics, requiring more data to learn patterns from raw data as opposed to hand-crafted features.
This finding highlights that the effectiveness of malware detectors is closely tied to the quality of the training data and the semantic richness it provides.
Furthermore, these findings suggest that carefully selected features that better capture malware semantics are crucial for effective detection, especially when training data is limited.
In contrast, models that learn directly from raw data often struggle to perform well under such data-scarce conditions.

\bulletpoint{APK feature}
Android malware detectors often leverage diverse features to enhance detection performance~\cite{arp2014drebin,kim2018multimodal,li2021robust}.
The reason is that each feature contributes to characterizing APKs, encoding their unique semantics.
One question arises: does the incorporation of more features necessarily enhance a detector's performance?
To investigate this, we remove individual features from the original feature set to assess the resulting performance.
For this experiment, it is essential that the required features of the chosen approaches are decomposable, allowing the sequential removal of individual features.
Accordingly, we spotlight Drebin, Kim et al., Xmal, and RAMDA, owing to their decomposable feature sets.
We exclude methods whose features are highly intertwined.
For instance, MamaDroid and MalScan rely on specific API calls to extract features from program graphs, making it challenging to remove individual features --- if API calls are removed, the graph features will also be removed.

\begin{table}[t]
    \centering
    \renewcommand{\arraystretch}{0.8}
    \caption{The impact of APK features on the effectiveness of Drebin, Kim et al., Xmal, and RAMDA.}
    \vspace{-0.1cm}
    \begin{adjustbox}{width=0.84\linewidth, center}
 \begin{tabular}{@{}l|cc|cc|cc|cc@{}}
 \toprule
 \multirow{2}{*}{\textbf{\raisebox{-0.2cm}{\begin{tabular}[c]{@{}l@{}}Feature\\ Combination\end{tabular}}}} & \multicolumn{2}{c|}{\textbf{Drebin}}      & \multicolumn{2}{c|}{\textbf{Kim et al.}}  & \multicolumn{2}{c|}{\textbf{Xmal}} & \multicolumn{2}{c}{\textbf{RAMDA}} \\ \cmidrule(l){2-9} 
 & \multicolumn{1}{c|}{\textbf{F1}} & \textbf{Acc.} & \multicolumn{1}{c|}{\textbf{F1}} & \textbf{Acc.} & \multicolumn{1}{c|}{\textbf{F1}} & \textbf{Acc.} & \multicolumn{1}{c|}{\textbf{F1}} & \textbf{Acc.} \\ \midrule
 Original   & \multicolumn{1}{c|}{\textbf{0.722}}& \textbf{0.948}  & \multicolumn{1}{c|}{0.726}& 0.944  & \multicolumn{1}{c|}{\textbf{0.698}}& \textbf{0.942}  & \multicolumn{1}{c|}{\textbf{0.636}}& \textbf{0.905}  \\ \midrule
 w/o hardware& \multicolumn{1}{c|}{0.717}& 0.947  & \multicolumn{1}{c|}{0.728}& 0.945  & \multicolumn{1}{c|}{N/A}  & N/A    & \multicolumn{1}{c|}{N/A}  & N/A    \\ \midrule
 w/o app-intent & \multicolumn{1}{c|}{0.622}& 0.936  & \multicolumn{1}{c|}{\textbf{0.776}}& \textbf{0.952}  & \multicolumn{1}{c|}{N/A}  & N/A    & \multicolumn{1}{c|}{0.428}& 0.845  \\ \midrule
 w/o permission & \multicolumn{1}{c|}{0.698}& 0.945  & \multicolumn{1}{c|}{0.692}& 0.939  & \multicolumn{1}{c|}{0.566}& 0.926  & \multicolumn{1}{c|}{0.526}& 0.882  \\ \midrule
 w/o api call& \multicolumn{1}{c|}{0.691}& 0.944  & \multicolumn{1}{c|}{0.699}& 0.942  & \multicolumn{1}{c|}{0.639}& 0.930  & \multicolumn{1}{c|}{0.482}& 0.880  \\ \midrule
 w/o opcode & \multicolumn{1}{c|}{N/A}  & N/A    & \multicolumn{1}{c|}{0.724}& 0.942  & \multicolumn{1}{c|}{N/A}  & N/A    & \multicolumn{1}{c|}{N/A}  & N/A    \\ \midrule
 w/o code string& \multicolumn{1}{c|}{0.702}& 0.944  & \multicolumn{1}{c|}{0.725}& 0.945  & \multicolumn{1}{c|}{N/A}  & N/A    & \multicolumn{1}{c|}{N/A}  & N/A    \\ \bottomrule
 \end{tabular}
    \end{adjustbox}
    \label{tab:combination_feats}
    \begin{tablenotes}[flushleft, labelsep=0]
 \footnotesize
 \item \noindent w/o means without. N/A denotes features that are not used in the original work.
 \end{tablenotes}
    \vspace{-0.3cm}
\end{table}

Table~\ref{tab:combination_feats} presents the outcomes of this experiment.
From the table, we observe that Drebin, Xmal, and RAMDA have performance degradation when one feature is removed.
This is intuitive, that each feature contains unique semantics describing the APKs and contributes to the overall effectiveness of a malware detector.
Interestingly, Kim et al. deviate from this trend.
In fact, even after omitting several features, its performance exceeds the original results.
For example, removing the app-intent features, the F1-score improves from 0.726 to 0.776, and Accuracy increases from 0.944 to 0.952.
These two observations underscore:
(i) each feature contains unique semantics describing the APKs and contributes to the overall effectiveness of a malware detector, and 
(ii) in some cases (\textit{e.g.,} specifc models or feature combination style), simply expanding the feature set does not guarantee enhanced performance.
This phenomenon underscores the importance of evaluating and justifying the inclusion of each feature in a malware detector based on the semantics.

We further examine the impact of feature combinations on the approach proposed by Kim et al. by varying the number of training epochs, in order to verify whether the observed phenomenon persists across different model configurations. Specifically, we set the training epochs to 50, 100, 150, and 200, and repeat the app-intent removal experiment under each setting to assess result consistency.
As shown in Table~\ref{tab:ablation_app_intent}, removing the app-intent feature consistently improves performance across all training epochs, in terms of both F1-score and accuracy. This demonstrates that the observed behavior is robust and not an artifact of a particular training configuration or convergence state.
These findings highlight an important insight: under certain conditions—such as specific classifiers or feature-combination strategies—the inclusion of additional features does not necessarily enhance detection performance and may even be detrimental. Consequently, researchers and practitioners should critically assess the contribution of individual features to overall performance, rather than assuming that incorporating more features will invariably yield better results.

\begin{table}[t]
\centering
\caption{Ablation study on the impact of app-intent information across different training epochs.}
\vspace{-0.1cm}
\label{tab:ablation_app_intent}
\begin{adjustbox}{width=0.6\linewidth, center}
\begin{tabular}{llcccc}
\toprule
\textbf{Method} & \textbf{Metric} & \textbf{50} & \textbf{100} & \textbf{150} & \textbf{200} \\
\midrule
\multirow{2}{*}{w/o app-intent}
& F-Score  & 0.740 & 0.776 & 0.753 & 0.753 \\
& Accuracy & 0.947 & 0.952 & 0.950 & 0.950 \\
\midrule
\multirow{2}{*}{original}
& F-Score  & 0.725 & 0.725 & 0.740 & 0.738 \\
& Accuracy & 0.944 & 0.944 & 0.952 & 0.942 \\
\bottomrule
\end{tabular}
\end{adjustbox}
\vspace{-0.5cm}
\end{table}

\bulletpoint{Model type}
Given specific features that describe app behaviors, different models can be used to extract semantics and detect malware.
The features reflect the richness of the available semantics, while the models determine how effectively these semantics are leveraged.
In this section, we select MamaDroid, HinDroid, and MalScan, which share the same feature set and feature representation, as shown in Table~\ref{tbl:systematization}, to evaluate the impact of different models on the effectiveness of malware detectors.
From the results in Table~\ref{tab:res_ratio} under scenario \ding{182}, we observe that HinDroid outperforms MamaDroid and MalScan.
This suggests that given a fixed feature set, different models demonstrate varying capabilities in extracting semantics and detecting Android malware.

\begin{table}[ht]
  \centering
  \caption{Performance of common ML models across API and permission feature subsets.}
  \vspace{-0.2cm}
  \label{tab:model_comparison}
  \begin{adjustbox}{width=0.87\linewidth}
    \begin{tabular}{lcccccc}
      \toprule
      \multirow{2}{*}{\textbf{Model}} &
        \multicolumn{2}{c}{\textbf{API+Permission}} &
        \multicolumn{2}{c}{\textbf{API}} &
        \multicolumn{2}{c}{\textbf{Permission}} \\
      \cmidrule(lr){2-3}\cmidrule(lr){4-5}\cmidrule(lr){6-7}
      & \textbf{Accuracy} & \textbf{F1-score} & \textbf{Accuracy} & \textbf{F1-score} & \textbf{Accuracy} & \textbf{F1-score} \\
      \midrule
      RandomForest & 0.984 & 0.929 & 0.975 & 0.897 & 0.973 & 0.868 \\
      SVM          & 0.978 & 0.897 & 0.975 & 0.897 & 0.962 & 0.794 \\
      DecisionTree & 0.984 & 0.927 & 0.965 & 0.857 & 0.959 & 0.819 \\
      KNN          & 0.973 & 0.865 & 0.975 & 0.897 & 0.954 & 0.761 \\
      MLP          & 0.981 & 0.916 & 0.973 & 0.886 & 0.973 & 0.872 \\
      \bottomrule
    \end{tabular}
  \end{adjustbox}
  \vspace{-0.4cm}
\end{table}

Beyond analyzing existing approaches, we conduct a small-scale empirical study to examine how different learning models perform given a fixed feature set. Specifically, we select two widely used feature types—API calls and permissions—to represent app behavior, and evaluate five representative models commonly adopted in Android malware detection: Random Forest, SVM, Decision Tree, KNN, and MLP.
For this experiment, we use the dataset shown in Figure~\ref{fig:new_data} and randomly sample 4,000 apps (3,600 benign and 400 malicious). The dataset is split into training, validation, and testing sets using an \texttt{8:1:1} ratio.

Intuitively, there is no universally optimal model for Android malware detection, as model performance is closely tied to the choice of feature representation. As shown in Table~\ref{tab:model_comparison}, when API calls are used as features, all evaluated classifiers achieve comparable performance. In contrast, when using permissions alone, MLP significantly outperforms the other models. When combining API calls and permission features, Random Forest achieves the best overall performance, with an accuracy of 98.4\% and an F1-score of 92.9\%.
Overall, our results suggest that ensemble-based models (\textit{e.g.}, Random Forest) and deep learning models (\textit{e.g.}, MLP) generally outperform other traditional classifiers (\textit{e.g.}, SVM, Decision Tree, and KNN) in Android malware detection tasks. Consequently, when designing new detectors, these models represent strong and practical baseline choices.

\begin{conclusionbox}
    \textit{Summary:}
    The effectiveness of ML-based malware detectors is shaped by various factors, such as APK features and model types. At the core of these influences is the richness of semantics embedded in the features and the models' capability to extract and utilize these semantics.
\end{conclusionbox}

\subsection{Robustness against real-world scenarios}
\label{robustness}
Previous works~\cite{pendlebury2019tesseract,li2023black,gao2024comprehensive} have started investigating the resilience of detectors when faced with real-world challenges such as malware evolution, obfuscation, and adversarial attacks.
Nonetheless, they often focus on a subset of these challenges or employ unrealistic experimental settings, potentially skewing their findings.
For instance, \textsc{Tesseract}~\cite{pendlebury2019tesseract} explores the impact of malware evolution, while Gao~\cite{gao2024comprehensive} assesses the robustness of detectors using an unrealistic balanced dataset.
This section aims to fill these gaps by thoroughly re-evaluating the robustness of the selected approaches in real-word settings, providing a clearer picture of where ML-based malware detection stands today.

\begin{figure*}[t]
    \centering
    \includegraphics[width=\linewidth]{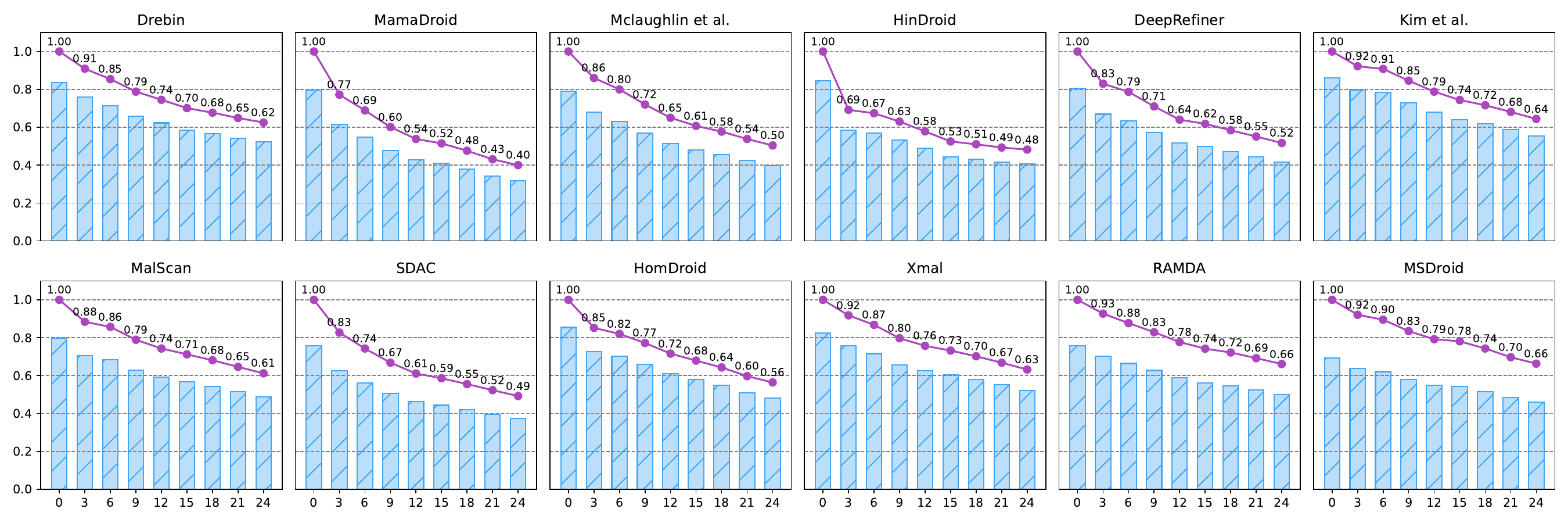}
    \caption{The performance of the selected techniques against diverse malware evolution periods. Columns display the absolute values of $AUT(F1, N)$.
    The line charts depict the relative percentage of $AUT(F1, N)$ against $AUT(F1, 0)$.}
    \label{fig:evolution}
\vspace{-0.5cm}
\end{figure*}

\bulletpoint{Malware evolution}
To quantify the impact of evolution on malware detectors, we utilize the $AUT(f, N)$~\cite{pendlebury2019tesseract}, where $f$ denotes the F1-score of a given approach, and $N$ represents the evolution period. 
The definition of $AUT(f, N)$ can be found in the Appendix~\ref{sub:aut}.
We set $N$ to 3, 6, 9, 12, 15, 18, 21, and 24 months in this experiment.
This metric ranges in (0, 1), where higher values indicate greater resilience to malware evolution.
In the study, we adopt a rolling algorithm over the data from 2011 to 2020 to calculate the $AUT(f, N)$.
Specifically, for each year, from 2011 to 2020, we first partition the data into training, validation, and testing sets with \texttt{8:1:1}.
Next, we train models with the training data, validate to get the best model, and evaluate it on the test set to get the F1-score as $AUT(F1, 0)$.
Then, the model is applied to test data in the next $N$ months, yielding $N$ F1-scores.
It is important to note that the $N$ months of test data may span across different years.
For example, if we use data from 2011 as the training set and set $N$ to 15 months, the corresponding test data will include apps collected from January 2012 to March 2013.
These scores are further used to calculate the $AUT(F1, N)$.
We repeat this process for each year from 2011 to 2020, resulting in ten distinct $AUT(F1, N)$ values for each $N$, if available.
For instance, when $N$ is 3 months, we can retrieve $AUT(F1, 3)$ values for each year from 2011 to 2019, since the test data for 3 months after training in 2020 is not available.
By averaging the $AUT(F1, N)$ values sourced from distinct yearly datasets, we chart the outcomes in Figure~\ref{fig:evolution}.

It is evident that malware evolution affects the effectiveness of the selected malware detectors.
This is expected: as malware evolves over time, its underlying semantics may change, making it more difficult for detectors to capture these behaviors.
Overall, most DL-based approaches exhibit stronger resilience to malware evolution than their TML-based counterparts.
Specifically, the majority of DL techniques retain around 60\% effectiveness even after two years, whereas many TML methods see their F1-scores drop to about 50\% over the same period.
This disparity can be attributed to the inherent capability of DL models to capture more intricate patterns that TML methods may miss.
Interestingly, McLaughlin et al. and DeepRefiner do not perform as well over time.
A plausible reason is their reliance on raw bytecode as input, which may contain substantial noise and complicate semantic extraction.
To enhance the robustness of ML-based Android malware detectors against malware evolution, two aspects are particularly important:
(i) extracting robust features that are less susceptible to temporal changes—for example, Xmal and RAMDA use carefully selected features (such as stable API calls and permissions) and tend to be more resilient to evolution;
and (ii) designing models that can effectively capture evolving malware semantics, such as persistent malicious behaviors—for instance, MsDroid employs a graph neural network to model structural relationships among code components, enabling it to better capture these evolving semantics.

\begin{figure*}[t]
    \centering
    \includegraphics[width=0.94\linewidth]{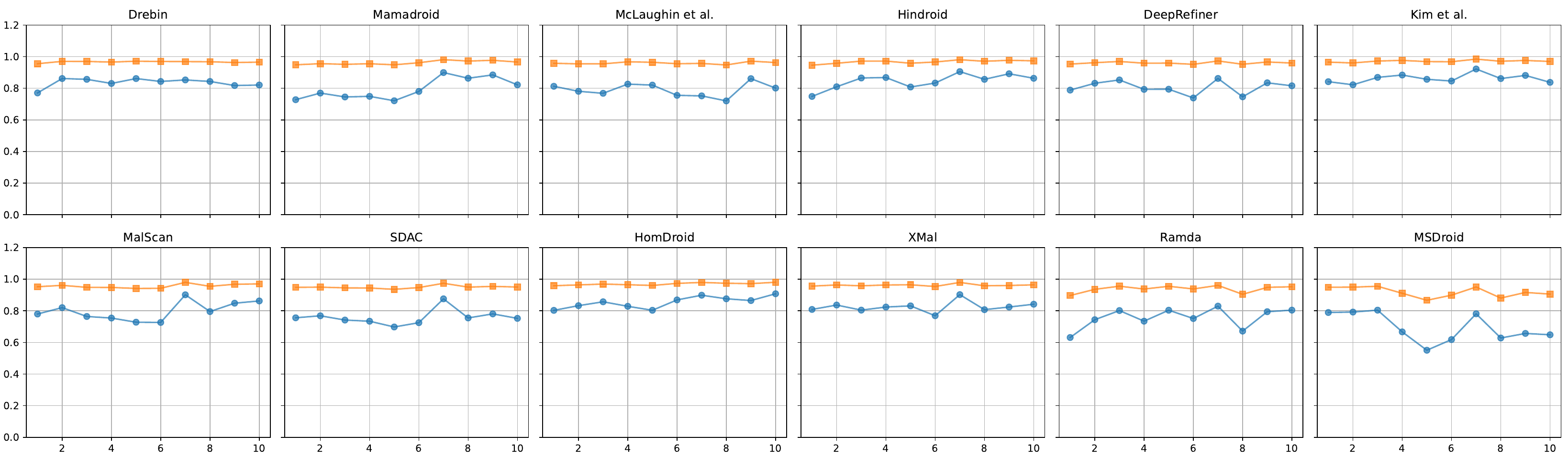}
    \vspace{-0.2cm}
    \caption{Performance of the selected detection approaches for each year from 2011 to 2020. Orange markers indicate Accuracy, and blue markers indicate F1-score.}
    \label{fig:stable_trend}
\vspace{-0.5cm}
\end{figure*}

\bulletpoint{Detector stability on different time period datasets}
Having examined the impact of malware evolution on detection effectiveness, we now turn to investigating whether the selected approaches can be applied to datasets collected in different time periods.
This experiment aims to assess the sensitivity of these approaches to temporal variations in the data. 
Such an analysis helps determine whether the approaches can be used reliably in practice and whether our findings are generalizable.
To cunduct this experiment, we partition the primary dataset into ten subsets based on the year of app collection, spanning from 2011 to 2020.
For each year-specific subset, we split the data into training (80\%), validation (10\%), and testing (10\%) sets.
Each approach is trained on the training set, with hyperparameters tuned using the validation set, and finally evaluated on the testing set.

Figure~\ref{fig:stable_trend} illustrates the performance of the selected approaches across different years.
We observe that most approaches maintain relatively stable performance over time, indicating that they can be effectively applied to datasets collected at different periods.
This stability suggests that the approaches are generally applicable rather than tailored to specific datasets, which in turn supports the claim that our findings are likely to generalize across different time periods.

\bulletpoint{Obfuscation}
Obfuscation techniques often alter an app's code. 
It can also affect the effectiveness of malware detectors~\cite{he2022msdroid}.
This part explores the influences of popular obfuscation techniques, \textit{i.e.,} renaming identifiers, encrypting resources, modifying code, and invoking reflection~\cite{aonzo2020obfuscapk}, on the selected approaches separately.
We apply the obfuscation strategies discussed earlier to the testing set in Table~\ref{tab:effectiveness_dataset}(\ding{182}).
Only the apps that can be successfully obfuscated by all the strategies are included in the obfuscated testing set, which contains 1,303 apps.
The effectiveness of the selected approaches, previously trained on the original training set, is then evaluated on this obfuscated testing set.

Table~\ref{table:res_obfuscation} summarizes the results.
Overall, most methods exhibit a decrease in performance when subjected to obfuscation.
This is expected, as obfuscation can hide malicious semantics, making them more difficult for detectors to capture. 
Specifically, the impact of obfuscation on detectors significantly depends on how the detectors utilize APK features.
For instance, Xmal and RAMDA show greater robustness against obfuscation.
A closer examination reveals that both approaches use API calls as features and rely on manually selected API calls as anchors, checking whether these APIs appear in the code.
We further observe that most of these anchor APIs are system APIs, which are less likely to be affected by the obfuscation techniques employed.
In other words, the obfuscation does not substantially modify the features used by these detectors.
In contrast, approaches such as MamaDroid and MsDroid, which rely heavily on code structure, tend to be more susceptible to code modification.
This suggests that the impact of obfuscation on detectors is closely related to the extent to which obfuscation techniques alter the semantics of the features they use.
Moreover, comparing MamaDroid and MsDroid—both structure-based—indicates that robustness to obfuscation is also influenced by the model's capability to extract semantics.
To better defend against obfuscation, it is therefore crucial to (i) select features that are less likely to be altered by obfuscation techniques and (ii) design models that can effectively extract the underlying semantics from these features.

\begin{table}[t]
    \centering
    \renewcommand{\arraystretch}{0.9}
    \caption{The F-score of the selected approaches under different obfuscation strategies~\cite{aonzo2020obfuscapk}.}
    \vspace{-0.1cm}
    \begin{adjustbox}{width=0.76\linewidth, center}
    \begin{tabular}{@{}c|c|c|c|c|c@{}}
    \toprule
    \diagbox[innerleftsep=-0.1em, innerrightsep=-0.3em, height=2.7em,width=7em]{\textbf{Approach}}{\textbf{Obfuscation}} & \textbf{\begin{tabular}[c]{@{}c@{}}Without \\ Obfus.\end{tabular}} & \textbf{\begin{tabular}[c]{@{}c@{}}Rename \\ Identifier\end{tabular}} & \textbf{\begin{tabular}[c]{@{}c@{}}Encrypt \\ Resource\end{tabular}} & \textbf{\begin{tabular}[c]{@{}c@{}}Modify \\ Code\end{tabular}} & \textbf{\begin{tabular}[c]{@{}c@{}}Reflect \\ Invocation\end{tabular}} \\ \midrule
    Drebin      & 0.732 & 0.702 & 0.701 & 0.732 & 0.732 \\ \midrule
    MamaDroid   & 0.653 & 0.274 & 0.449 & 0.150 & 0.461 \\ \midrule
    Mclaughlin et al.  & 0.750 & 0.699 & 0.722 & 0.175 & 0.727 \\ \midrule
    HinDroid    & 0.750 & 0.750 & 0.741 & 0.750 & 0.735 \\ \midrule
    DeepRefiner & 0.692 & 0.618 & 0.658 & 0.297 & 0.612 \\ \midrule
    Kim et al.  & 0.795 & 0.795 & 0.611 & 0.795 & 0.798 \\ \midrule
    MalScan     & 0.675 & 0.675 & 0.675 & 0.681 & 0.687 \\ \midrule
    SDAC & 0.563 & 0.538 & 0.552 & 0.495 & 0.552 \\ \midrule
    HomDroid    & 0.729 & 0.751 & 0.706 & 0.728 & 0.739 \\ \midrule
    Xmal & 0.727 & 0.727 & 0.727 & 0.727 & 0.727 \\ \midrule
    RAMDA& 0.635 & 0.635 & 0.635 & 0.635 & 0.635 \\ \midrule
    MSDroid     & 0.672 & 0.675 & 0.610 & 0.356 & 0.494 \\ \bottomrule
    \end{tabular}
    \end{adjustbox}
    \label{table:res_obfuscation}
    \vspace{-0.4cm}
\end{table}

\bulletpoint{Adversarial attack}
We now utilize the dataset from Table~\ref{tab:effectiveness_dataset}(\ding{182}) to explore the impact of adversarial attacks on the performance of these selected approaches.
It is worth noting that we exclude approaches Mclaughlin et al. and DeepRefiner from this experiment.
This is because they truncate features at a certain size, making them easily bypassable.
Attackers can embed malicious code in the ignored part to evade detection.

To investigate the impact of adversarial attacks, we employ two primary strategies: Jacobian Saliency Map Attack (JSMA) and Randomized Input (RI).
These two techniques are chosen for their simplicity and effectiveness, compromising most ML-based malware detectors, as shown in Table~\ref{tab:adversarial}.
This suggests that more advanced attack strategies could pose even greater threats to these detectors~\cite{he2023efficient,zhao2021structural}.
For JSMA, we first train a substitute model on the training set, and then apply JSMA to generate adversarial samples. 
Specifically, we utilize a Multi-Layer Perceptron (MLP) as the substitute model for conventional ML-based classifiers (e.g., SVM, KNN, and Random Forest), as MLP can effectively approximate the decision boundaries learned by these classifiers while remaining fully differentiable for gradient-based attacks such as JSMA.
For GNN-based detectors (e.g., MsDroid), we adopt a Graph Convolutional Network (GCN) as the substitute model, as GCN captures both node features and edge structures through message passing, and provides a simple yet representative architecture for modeling graph-based malware detectors.
During the crafting of adversarial samples, we calculate the Jacobian matrix to identify the most influential features and iteratively modify them until the samples are misclassified by the target model or a maximum number of modifications is reached.
For RI, we randomly change a certain percentage of features in the testing set to produce adversarial examples.
When crafting adversarial samples, we follow~\cite{li2023black,chen2019android} to ensure the feature modifications are domain-mappable and can be repackaged to APKs.
For feature-vector-based approaches like Drebin, we follow~\cite{chen2019android} by constraining modifications to vector bits that are 0, changing them to 1.
This ensures that all required permissions and functions remain unchanged, keeping the app's functionality unaffected.
For graph-based solutions, we introduce non-disruptive modifications to the graph structures that do not alter the app's functionality.
The graph structure includes non-leaf nodes (user-defined functions) and leaf nodes (Android API). 
When adding a new edge, we select a non-leaf node in the graph and add a try block with a callee chosen from leaf nodes.
Since leaf nodes do not call any other nodes, the added edge does not affect the app's functionality.
To measure the robustness against adversarial attacks, we use two metrics from~\cite{li2023black}: Adversarial Success Rate (ASR) and Adversarial Perturbation Ratio (APR).
The detailed definitions can be found in the Appendix~\ref{sub:aut}.
Both metrics range (0, 1).
A higher ASR indicates increased vulnerability to adversarial attacks, while a higher APR suggests greater robustness against such attacks due to the model's ability to withstand perturbations.

By analyzing Table~\ref{tab:adversarial}, we note that JSMA yields an average ASR of 98.8\% across the evaluated approaches, indicating their vulnerability to such basic adversarial attacks, let alone more sophisticated ones~\cite{he2023efficient}.
The employed strategies are less effective on MamaDroid, HomDroid, and RAMDA.
The reason for MamaDroid and HomDroid could be linked to their abstraction of APKs' graph structures.
Interestingly, MamaDroid, HomDroid, and MalScan all utilize API calls and program graphs; the former two methods demonstrate greater resilience.
The key difference lies in their handling of program graphs --- MamaDroid abstracts these graphs using family and package names, and HomDroid employs social network triads for abstraction, in contrast to MalScan's direct use of the graphs.
These abstractions make the malicious semantics encoded in graph structures more resistant to perturbation, enhancing the robustness of the detectors built on them.
Additionally, RAMDA's resilience can be attributed to its customized Autoencoder, which learns compressed representations of benign apps, making it more defensive to adversarial examples.
These findings suggest that abstracting feature representations or augmenting models' defensive capabilities could improve malware detectors' robustness to adversarial attacks.

\begin{table}[t]
    \renewcommand{\arraystretch}{0.73}
    \caption{The robustness of our selected approaches on various adversarial attacks.}
    \vspace{-0.2cm}
    \begin{adjustbox}{width=0.78\linewidth, center}
 \begin{tabular}{@{}c|cc|c||c|cc|c@{}}
  \toprule
  \multirow{2}{*}{\textbf{\makecell[c]{Selected\\ Approach}}} & \multicolumn{2}{c|}{\textbf{ASR}}      & \multirow{2}{*}{\textbf{APR}} & \multirow{2}{*}{\textbf{\makecell[c]{Selected\\ Approach}}} & \multicolumn{2}{c|}{\textbf{ASR}}      & \multirow{2}{*}{\textbf{APR}} \\ \cmidrule(lr){2-3} \cmidrule(lr){6-7}
   & \multicolumn{1}{c|}{\textbf{JSMA}} & \textbf{RI} &    &  & \multicolumn{1}{c|}{\textbf{JSMA}} & \textbf{RI} &    \\ \midrule
  Drebin  & \multicolumn{1}{c|}{1.000}  & 0.237& 0.001 & SDAC    & \multicolumn{1}{c|}{1.000}  & 0.146& 0.001 \\ \midrule
  MamaDroid  & \multicolumn{1}{c|}{0.972}  & 0.187& 0.021 & HomDroid& \multicolumn{1}{c|}{0.979}  & 0.303& 0.324 \\ \midrule
  HinDroid& \multicolumn{1}{c|}{1.000}  & 0.068& 0.004 & Xmal    & \multicolumn{1}{c|}{1.000}  & 0.142& 0.013 \\ \midrule
  Kim et al.  & \multicolumn{1}{c|}{1.000}  & 0.000& 0.004 & RAMDA   & \multicolumn{1}{c|}{0.931}  & 0.300& 0.023 \\ \midrule
  MalScan & \multicolumn{1}{c|}{1.000}  & 0.788& 0.001 & MSDroid & \multicolumn{1}{c|}{1.000}  & 0.022& 0.013 \\ \bottomrule
  \end{tabular}
    \end{adjustbox}
    \label{tab:adversarial}
\vspace{-0.4cm}
\end{table}

\begin{conclusionbox}
    \textit{Summary:}
    ML-based malware detectors are inherently susceptible to real-world challenges such as malware evolution, code obfuscation, and adversarial attacks. Our findings suggest that detectors capable of capturing stable features and underlying malware semantics are more robust to these challenges. These observations highlight the need for future research to focus on identifying, modeling, and integrating semantically meaningful behavioral representations that remain invariant under transformation and evolution.
\end{conclusionbox}

\subsection{Efficiency}
\label{efficiency}
In this section, we scrutinize the efficiency of the selected methods across datasets from different years, 
focusing on the end-to-end efficiency of the entire detection pipeline, including feature extraction, transformation, and ML modeling.
For feature extraction, we utilize average time for processing one APK and memory overhead during the extraction process.
For feature transformation, we measure the time taken to retrieve and encode the required features from pre-extracted features.
We normalize the time based on processing 20,000 apps per year.
For ML modeling, we evaluate the time taken for model training and prediction.
All experiments are performed on a server with 32-core CPUs operating at 2.10 GHz, 251 GB of physical memory, and two GPUs, each with 32 GB of memory.

\begin{figure}[t]
    \centering
    \begin{subfigure}{0.46\linewidth}
        \centering
        \includegraphics[width=\linewidth]{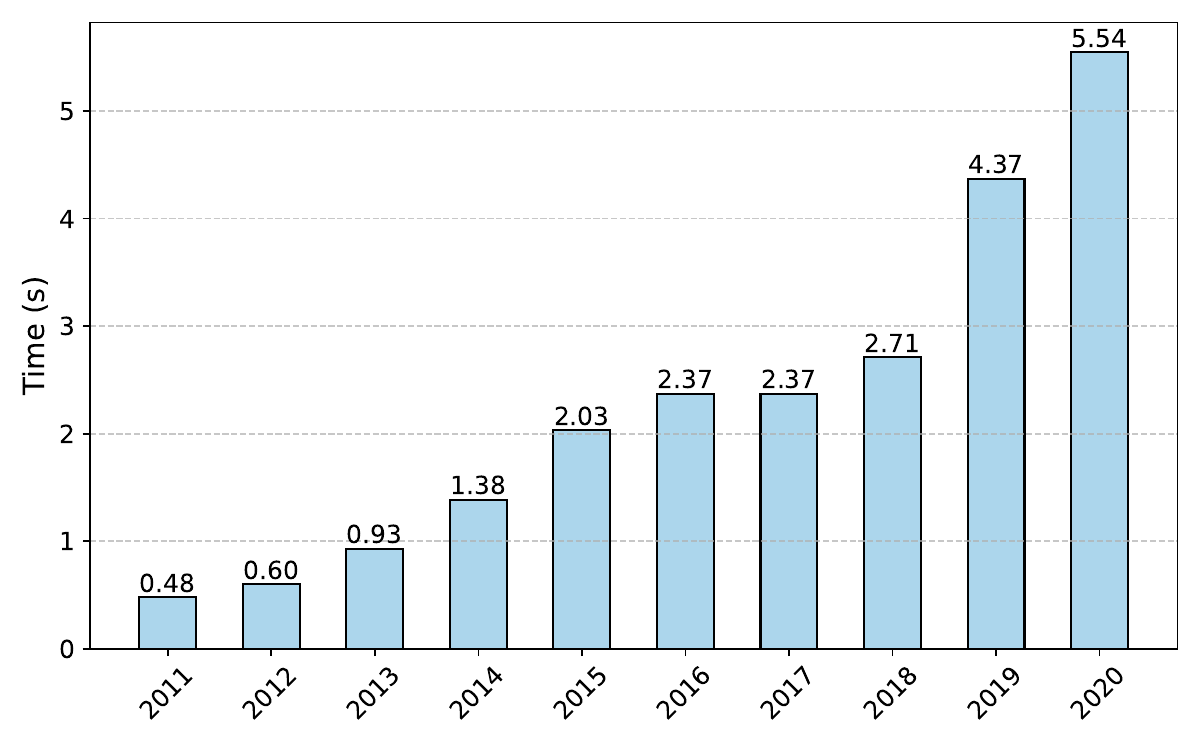}
        \caption{Time efficiency}
        \label{fig:efficiency_time}
    \end{subfigure}
    \begin{subfigure}{0.46\linewidth}
        \centering
        \includegraphics[width=\linewidth]{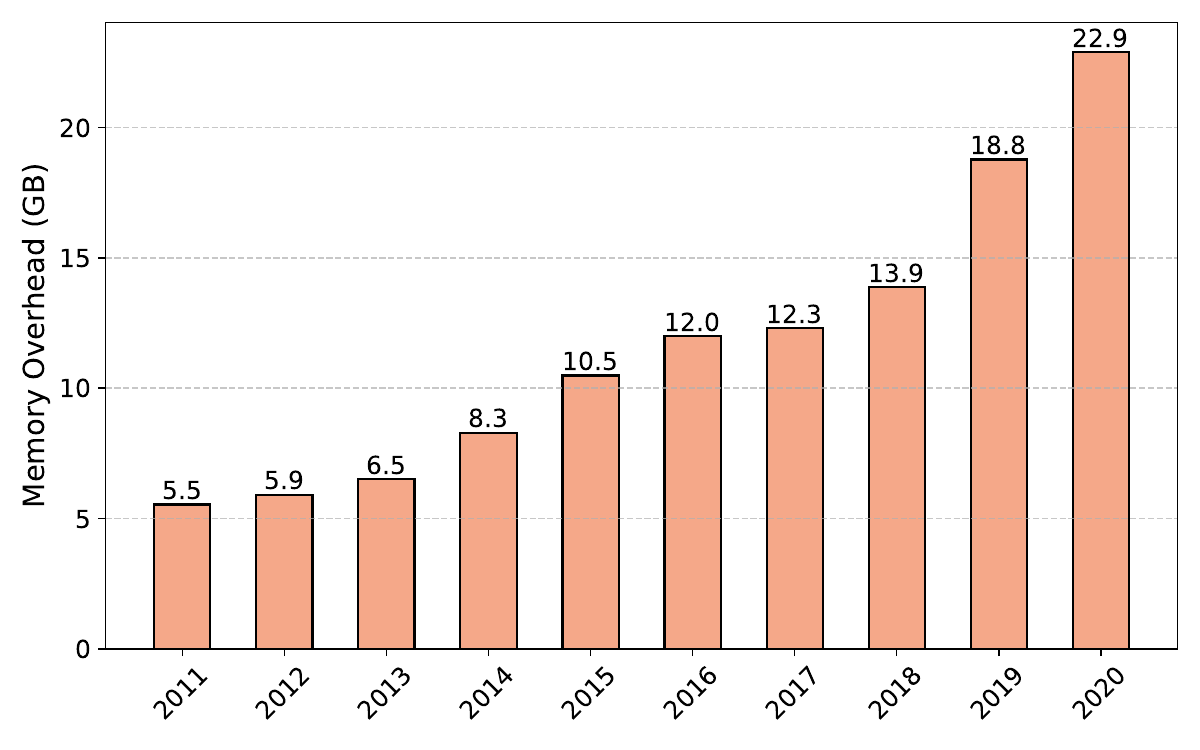}
        \caption{Memory efficiency}
        \label{fig:efficiency_memory}
    \end{subfigure}

    \caption{The efficiency of extracting features from APKs in terms of time and memory.}
    \label{fig:efficiency_combined}
    \vspace{-0.3cm}
\end{figure}

\begin{figure}[t]
    \centering
    \includegraphics[width=0.56\linewidth]{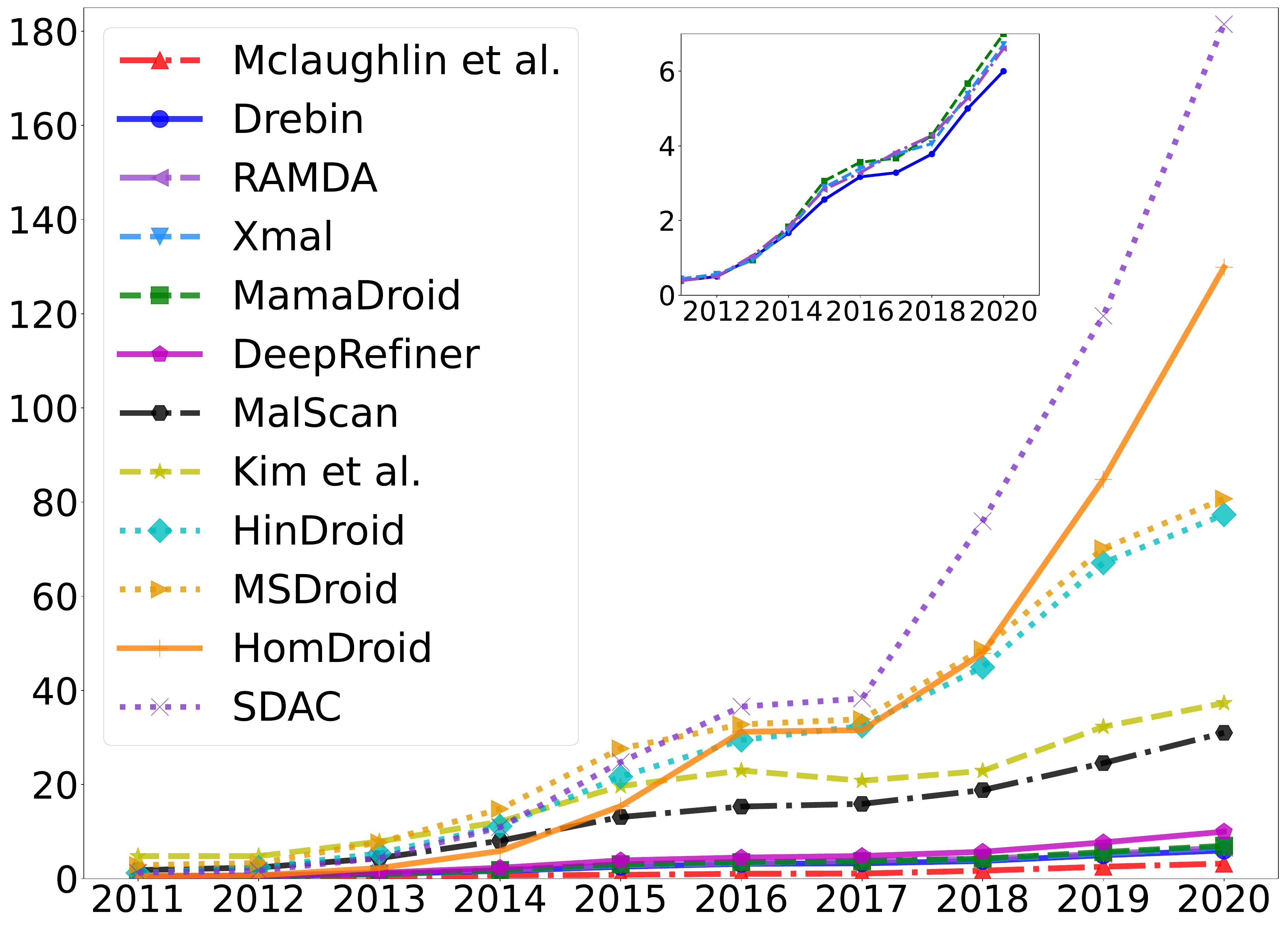}
    \vspace{-0.2cm}
    \caption{The efficiency of feature transformation of the selected approaches. The x-axis shows the dataset year, and the y-axis indicates the utilized time in hours.}
    \label{fig:efficiency_encoding}
    \vspace{-0.5cm}
\end{figure}
\bulletpoint{Feature extraction}
We first measure the performance overhead incurred during feature extraction from APKs.
The extraction process supports multi-process execution to enable concurrent feature extraction; in our experiments, we set the number of processes to 16.
We record the average time required to extract features from a single APK, as well as the average memory consumption.
Memory usage is monitored across all processes at 30-second intervals, and the mean value is reported as the memory overhead.

Figure~\ref{fig:efficiency_combined} illustrates the time and memory efficiency of feature extraction. We observe that both time and memory overheads increase over the years, a trend that can be attributed to the growing complexity and size of APKs, which require more resources for effective feature extraction. In general, time and memory consumption are positively correlated, with higher extraction times typically accompanied by higher memory usage.
Despite this upward trend, the observed overheads remain acceptable for practical use of \codename. For example, extracting features from an APK released in 2020 takes approximately 5.5 seconds, with an overall memory consumption of about 22.9~GB when using 16 concurrent processes --- well within the capabilities of modern systems.

\begin{table*}[t]
    \centering
    \renewcommand{\arraystretch}{1.1}
    \caption{The efficiency of selected approaches across datasets from different years, covering training and testing phases.}
    \begin{adjustbox}{width=0.98\textwidth}
    \begin{tabular}{@{}cclc|c|c|c|c|c|c|c|c|c|c|c@{}}
    \cmidrule(l){4-15}
    \multicolumn{3}{c}{}   & Drebin & MamaDroid & Mclaughlin et al. & HinDroid & DeepRefiner & Kim et al. & MalScan & SDAC & HomDroid & Xmal   & RAMDA & MSDroid \\ \midrule
    \multicolumn{1}{c|}{\multirow{10}{*}{\raisebox{-2.8cm}{\rotatebox{90}{\textbf{\begin{tabular}[c]{@{}c@{}}Training\end{tabular}}}}}} & \multicolumn{2}{c|}{\textbf{2011}} & 15.11s  & 1.90s & 128m22s & 2m19s & 369m18s & 98m47s & 1m14s & 84m47s & 2.00s & 1m15s  & 1m6s & 20m19s \\ \cmidrule(l){2-15} 
    \multicolumn{1}{c|}{} & \multicolumn{2}{c|}{\textbf{2012}} & 20.88s  & 1.94s & 97m01s  & 2m18s & 359m35s   & 92m47s  & 1m18s   & 89m05s   & 2.04s     & 1m25s  & 1m11s & 37m54s \\ \cmidrule(l){2-15} 
    \multicolumn{1}{c|}{} & \multicolumn{2}{c|}{\textbf{2013}} & 22.37s  & 2.12s & 172m20s & 2m13s & 477m18s   & 115m40s & 1m24s   & 115m35s  & 2.04s     & 50.36s & 1m06s & 40m11s \\ \cmidrule(l){2-15} 
    \multicolumn{1}{c|}{} & \multicolumn{2}{c|}{\textbf{2014}} & 32.49s  & 2.46s & 128m13s & 2m40s & 629m26s   & 104m03s & 1m33s   & 132m55s  & 2.20s     & 2m20s  & 1m17s & 42m07s \\ \cmidrule(l){2-15} 
    \multicolumn{1}{c|}{} & \multicolumn{2}{c|}{\textbf{2015}} & 26.40s  & 2.60s & 411m03s & 2m44s & 730m41s   & 115m27s & 1m38s   & 250m07s  & 2.22s     & 2m10s  & 1m20s & 42m52s \\ \cmidrule(l){2-15} 
    \multicolumn{1}{c|}{} & \multicolumn{2}{c|}{\textbf{2016}} & 32.62s  & 2.99s & 234m19s & 2m51s & 867m25s   & 135m50s & 1m42s   & 470m38s  & 2.26s     & 1m05s  & 1m19s & 142m23s \\ \cmidrule(l){2-15} 
    \multicolumn{1}{c|}{} & \multicolumn{2}{c|}{\textbf{2017}} & 22.36s  & 2.75s & 192m41s & 2m28s & 997m04s   & 145m24s & 1m37s   & 296m05s  & 2.22s     & 1m54s  & 1m20s & 47m22s \\ \cmidrule(l){2-15} 
    \multicolumn{1}{c|}{} & \multicolumn{2}{c|}{\textbf{2018}} & 19.97s  & 2.91s & 227m48s & 2m36s & 933m12s   & 128m32s & 1m34s   & 716m43s  & 2.21s     & 1m19s  & 1m21s & 155m50s \\ \cmidrule(l){2-15} 
    \multicolumn{1}{c|}{} & \multicolumn{2}{c|}{\textbf{2019}} & 26.83s  & 3.03s & 514m18s & 2m34s & 968m50s   & 185m43s & 1m32s   & 918m09s  & 2.22s     & 1m34s  & 1m18s & 130m41s \\ \cmidrule(l){2-15} 
    \multicolumn{1}{c|}{} & \multicolumn{2}{c|}{\textbf{2020}} & 22.87s  & 2.69s & 771m44s & 2m22s & 1064m30s  & 148m08s & 1m29s   & 867m55s  & 2.17s     & 2m09s  & 1m12s & 93m08s \\ \midrule \midrule
    \multicolumn{1}{c|}{\multirow{10}{*}{\raisebox{-2cm}{\rotatebox{90}{\textbf{\begin{tabular}[c]{@{}c@{}}Testing\end{tabular}}}}}}  & \multicolumn{2}{c|}{\textbf{2011}} & 0.98s   & 0.06s & 16.03s  & 29.82s & 42.04s & 0.62s & 18.02s   & 54.43s & 1.20s     & 0.32s   & 0.07s  & 4.33s    \\ \cmidrule(l){2-15} 
    \multicolumn{1}{c|}{} & \multicolumn{2}{c|}{\textbf{2012}} & 1.36s   & 0.06s & 16.94s  & 30.40s & 45.50s & 0.89s & 22.23s   & 53.97s  & 1.30s  & 0.32s   & 0.07s  & 6.15s   \\ \cmidrule(l){2-15} 
    \multicolumn{1}{c|}{} & \multicolumn{2}{c|}{\textbf{2013}} & 1.48s   & 0.06s & 16.25s  & 31.53s & 50.36s & 1.10s & 19.27s   & 1m21s   & 1.22s  & 0.34s   & 0.07s  & 7.14s   \\ \cmidrule(l){2-15} 
    \multicolumn{1}{c|}{} & \multicolumn{2}{c|}{\textbf{2014}} & 1.53s   & 0.06s & 24.69s  & 33.10s & 59.22s & 1.11s & 24.83s   & 1m51s   & 1.36s  & 0.32s   & 0.09s  & 11.29s  \\ \cmidrule(l){2-15} 
    \multicolumn{1}{c|}{} & \multicolumn{2}{c|}{\textbf{2015}} & 1.73s   & 0.06s & 28.63s  & 34.95s & 1m12s  & 0.97s & 23.59s   & 2m54s   & 1.41s  & 0.39s   & 0.07s  & 10.92s  \\ \cmidrule(l){2-15} 
    \multicolumn{1}{c|}{} & \multicolumn{2}{c|}{\textbf{2016}} & 2.16s   & 0.06s & 40.04s  & 36.50s & 1m23s  & 1.19s & 23.92s   & 3m37s   & 1.54s  & 0.33s   & 0.08s  & 18.72s  \\ \cmidrule(l){2-15} 
    \multicolumn{1}{c|}{} & \multicolumn{2}{c|}{\textbf{2017}} & 1.46s   & 0.06s & 34.05s  & 35.25s & 1m12s  & 0.92s & 26.19s   & 4m16s   & 1.50s  & 0.31s   & 0.09s  & 17.96s  \\ \cmidrule(l){2-15} 
    \multicolumn{1}{c|}{} & \multicolumn{2}{c|}{\textbf{2018}} & 1.28s   & 0.06s & 36.39s  & 35.17s & 1m18s  & 0.89s & 22.67s   & 6m21s   & 1.28s  & 0.32s   & 0.09s  & 17.96s  \\ \cmidrule(l){2-15} 
    \multicolumn{1}{c|}{} & \multicolumn{2}{c|}{\textbf{2019}} & 1.74s   & 0.06s & 57.58s  & 34.93s & 1m33s  & 1.18s & 24.43s   & 8m55s   & 1.45s  & 0.33s   & 0.09s  & 21.14s  \\ \cmidrule(l){2-15} 
    \multicolumn{1}{c|}{} & \multicolumn{2}{c|}{\textbf{2020}} & 1.40s   & 0.06s & 1m12s   & 33.90s & 1m49s  & 0.91s & 23.62s   & 8m27s   & 1.52s  & 0.41s   & 0.07s  & 19.54s  \\ \bottomrule
    \end{tabular}
    \end{adjustbox}
    \label{tab:efficiency_ml_modeling}
\vspace{-0.3cm}
\end{table*}

\bulletpoint{Feature transformation}
With pre-extracted features available, we further evaluate the time efficiency of the selected methods in retrieving and encoding their required features. All feature transformation procedures are executed using 16 concurrent processes to ensure consistency across methods.

Figure~\ref{fig:efficiency_encoding} presents the time overhead incurred by different approaches during feature transformation. We observe a clear increasing trend over time, with more recent APKs requiring longer processing times. This trend is consistent with the rapid growth in both the size and structural complexity of Android applications, which imposes higher computational costs during feature encoding. These results highlight the importance of designing scalable and efficient feature encoding strategies to keep pace with the evolution of Android ecosystems.
Notably, substantial variance in time consumption is observed across different detection approaches. This variation aligns with our expectations, as different methods rely on distinct feature types and encoding mechanisms with varying computational complexity. For example, SDAC exhibits the highest time overhead due to its need to recursively traverse the program graph to generate API call sequences, an operation that scales poorly with increasing code complexity. In contrast, approaches such as Xmal and RAMDA are considerably more time-efficient, as they rely on a single traversal of the program graph to extract API calls, resulting in significantly lower processing overhead.

Overall, these findings demonstrate that the choice of feature representation and encoding strategy has a substantial impact on time efficiency, underscoring a key trade-off between expressive feature modeling and computational scalability in ML-based Android malware detection.

\bulletpoint{ML modeling}
For each yearly dataset, we split the samples into training, validation, and testing sets using an \texttt{8:1:1} ratio. We then measure the time required by the selected approaches for both model training and testing. For deep learning (DL)-based methods, we adopt an early-stopping strategy during training to prevent overfitting and ensure fair comparison.

Table~\ref{tab:efficiency_ml_modeling} reports the training and testing time of the evaluated approaches.
From a cross-method perspective, we observe that most DL-based techniques incur substantially higher training costs than traditional machine learning (TML) methods.
Moreover, training time is strongly correlated with model complexity, with deeper or more expressive architectures requiring longer training durations.
From a longitudinal perspective, training time consistently increases over the years, reflecting the growing complexity of Android applications and the expanding dimensionality of extracted features.
When considered jointly with detection effectiveness (Section~\ref{effectiveness}), these results reveal that higher resource consumption does not necessarily translate into superior detection performance.
This observation highlights a fundamental trade-off between effectiveness and efficiency in ML-based Android malware detection.
Designing practical detectors therefore requires careful consideration of this balance, where incorporating semantically meaningful features plays a critical role in achieving strong performance without excessive computational overhead.
During the testing phase, we observe trends similar to those in training, with prediction time increasing modestly over time.
Nevertheless, most approaches are able to complete inference within seconds when evaluating approximately 2,000 apps, indicating that they remain suitable for near real-time malware detection in practical deployment scenarios.

\begin{conclusionbox}
    \textit{Summary:}
    The basic costs of feature extraction and malware prediction for ML-based detectors remain within acceptable limits for practical deployment. However, the primary efficiency bottleneck lies in feature transformation, where both what features are selected and how they are represented have a significant impact on encoding time. Notably, detector efficiency does not necessarily correlate with detection effectiveness, underscoring the importance of balancing effectiveness and efficiency.
    In particular, features should be well aligned with the capabilities of the chosen model to achieve an optimal trade-off. For example, when employing classifiers such as SVMs, extracting complex graph-based features may be unnecessary, as such models are not well-suited to fully exploit the graph structure. Future work should focus on principled feature-model co-design, enabling efficient yet semantically expressive detection pipelines.
\end{conclusionbox}

\subsection{Further Experimental Validation}
\label{further_experiments}
\begin{table}[t]
\centering
\caption{Test Set Performance Comparison: APKTool vs Androguard Features}
\label{tab:results}
\renewcommand{\arraystretch}{1.1}
\begin{adjustbox}{width=0.83\linewidth}
\begin{tabular}{l|cccc|cccc}
\hline
\multirow{2}{*}{\textbf{Model}} & \multicolumn{4}{c|}{\textbf{APKTool}} & \multicolumn{4}{c}{\textbf{Androguard}} \\
\cline{2-9}
& \textbf{Acc.} & \textbf{Prec.} & \textbf{Rec.} & \textbf{F1} & \textbf{Acc.} & \textbf{Prec.} & \textbf{Rec.} & \textbf{F1} \\
\hline
RandomForest & 0.975 & 0.941 & 0.800 & 0.865 & 0.984 & 0.886 & 0.975 & 0.929 \\
SVM          & 0.963 & 0.931 & 0.675 & 0.783 & 0.978 & 0.921 & 0.875 & 0.897 \\
DecisionTree & 0.960 & 0.786 & 0.825 & 0.805 & 0.984 & 0.905 & 0.950 & 0.927 \\
KNN          & 0.960 & 0.900 & 0.675 & 0.771 & 0.973 & 0.941 & 0.800 & 0.865 \\
MLP          & 0.973 & 0.939 & 0.775 & 0.849 & 0.981 & 0.884 & 0.950 & 0.916 \\
\hline
\end{tabular}
\end{adjustbox}
\vspace{-0.5cm}
\end{table}

\subsubsection{\textbf{Influence of Reverse Engineering Tools}}
\label{tool_influence}
As discussed in Section~\ref{sub:compare}, different reverse engineering tools may extract divergent features from the same APK, potentially affecting the performance of ML-based malware detectors.
To systematically investigate this effect, we randomly select 4,000 apps from the additional dataset shown in Figure~\ref{fig:new_data}, maintaining a malware ratio of 10\%.
We then extract features using two widely adopted reverse engineering tools: Androguard~\cite{androguard} and APKTool~\cite{apkt}.
In this experiment, we focus on two commonly used feature types—API calls and permissions.
We split the dataset into training, validation, and testing sets, and train five representative ML models --- Random Forest, SVM, Decision Tree, KNN, and MLP --- using features extracted by each tool independently.
Each trained model is evaluated on its corresponding test set, and the results are compared to assess the impact of tool choice on detection performance.

Our analysis reveals that the features extracted by the two tools are not always consistent.
In some cases, certain API calls or permissions are identified by one tool but missed by the other.
For instance, for the APK \texttt{00002EA4***0B6FB1}, Androguard extracts 15 additional API calls, such as \texttt{getRunningAppProcesses} and \texttt{getRunningTasks}, that are not detected by APKTool.
Such discrepancies are expected, as different tools employ distinct parsing strategies and static analysis techniques, leading to variations in feature coverage.
We further evaluate the performance of models trained on features extracted by each tool, with the results summarized in Table~\ref{tab:results}.
The results show noticeable performance differences across models depending on the tool used for feature extraction, demonstrating that the choice of reverse engineering tool can significantly influence experimental outcomes
This finding underscores the importance of standardizing the feature extraction toolchain when conducting comparative evaluations.
Accordingly, in this work, we use the same tool for all approaches to ensure fair and reproducible comparisons.
Given its consistently stronger performance across most settings, we select Androguard as the default feature extraction tool in \codename.

\subsubsection{\textbf{Detection Effectiveness}}
\label{sec:detect_effectiveness}
As described in Section~\ref{sec:dataset}, we use the primary dataset spanning 2011-2020 as the main benchmark for evaluating the effectiveness of the selected representative approaches, providing a comprehensive view of the state of ML-based Android malware detection.
To further validate the generalizability of our findings and examine whether they remain valid on more recent data, we conduct an additional evaluation using a supplementary dataset collected between 2021 and 2024, as illustrated in Figure~\ref{fig:new_data}.
For consistency, we split the dataset into training, validation, and testing sets using an \texttt{8:1:1} ratio, while maintaining a goodware-to-malware ratio of 9:1, following the same experimental setup as in Section~\ref{effectiveness}.

Table~\ref{tab:2021_2024_results} reports the F1-score and accuracy of the selected approaches on the additional datasets.
We observe trends consistent with those obtained from the primary dataset: under realistic settings, Hindroid, Kim et al., and Homdroid consistently outperform other methods and achieve comparable F1-score and accuracy.
These results indicate that traditional machine learning (TML)-based and deep learning (DL)-based approaches continue to exhibit similar effectiveness on recent data, further supporting our earlier finding that DL-based methods may require a larger volume of malware samples to effectively distill malicious semantics and surpass TML-based approaches.

This experiment serves as a lightweight yet fundamental validation that the selected representative approaches—and our key observations—remain effective on more recent datasets.
Other aspects of detector behavior, such as robustness and efficiency, are less sensitive to dataset time periods and therefore are not re-evaluated on the additional dataset.
Further discussion on the influence of dataset temporal characteristics is provided in Section~\ref{sec:discussion}.

\begin{table}[t]
\centering
\caption{Average F1-Score and Accuracy for different methods on the additional datasets from 2021 to 2024.}
\begin{adjustbox}{width=0.8\linewidth}
\begin{tabular}{lcccccc}
	\toprule
	\textbf{Method} & Drebin & Mamadroid & Mclaughin et al. & Hindroid & DeepRefiner & Kim et al. \\
\midrule
	\textbf{F1-Score} & 0.857 & 0.716 & 0.717 & 0.900 & 0.707 & 0.879 \\
	\textbf{Accuracy} & 0.973 & 0.744 & 0.744 & 0.981 & 0.742 & 0.977 \\
\midrule
\midrule
	\textbf{Method} & Malscan & SDAC & Homdroid & Xmal & RAMDA & MsDroid \\
\midrule
	\textbf{F1-Score} & 0.716 & 0.735 & 0.888 & 0.670 & 0.723 & 0.823 \\
	\textbf{Accuracy} & 0.743 & 0.747 & 0.979 & 0.734 & 0.932 & 0.958 \\
\bottomrule
\end{tabular}
\end{adjustbox}
\label{tab:2021_2024_results}
\vspace{-0.3cm}
\end{table}

\section{Findings and Recommendations}
\label{sec:findings}
We now draw from our findings to discuss the current state of ML-based Android malware detection and put forth recommendations to guide future research in this area.

\bulletpoint{Findings}
We summarize our key findings as follows:

\noindent \textit{(1) Current ML-based Android malware detectors still face open challenges.}
While ML models have been evidently effective in detecting malware~\cite{he2022msdroid,kim2018multimodal,li2021robust,mariconti2016mamadroid}, their effectiveness is still far from satisfactory when faced with challenging scenarios such as limited data size, rapid malware evolution, and adversarial attacks.

\noindent \textit{(2) Selecting features relevant to malware semantics is crucial for effective detection.}
A wide range of APK features, such as permissions and intents, have been leveraged for malware detection~\cite{xu2018deeprefiner,arp2014drebin,kim2018multimodal}.
These features serve to profile app behavior and play a critical role in determining detector effectiveness.
Our findings suggest that heuristic feature selection can effectively reduce noise in the feature space, allowing models to more readily identify malicious patterns --- particularly in data-scarce settings.
However, we also observe that the semantics captured by different feature types are not always complementary, and principled strategies for combining them remain underexplored.
Consequently, naively incorporating additional features does not necessarily yield improved detection performance.

\noindent \textit{(3) More complex models are not a silver bullet in designing malware detectors.}
The literature reflects the trend towards employing more powerful ML models to detect malware~\cite{xu2018deeprefiner,arp2014drebin}.
Our analysis reveals that DL-based methods appear to be more resilient than TML-based approaches in adapting to malware evolution.
However, we also find that DL-based approaches are less effective than TML-based methods when the malware-to-goodware ratio is low.
This suggests that utilizing more complex models is not a universal solution for malware detection.
Instead, the choice of models should take into account the richness of the semantic information derived from the features.
One more practical observation is given feature sets including discrete features (\textit{i.e.,} permissions and API calls), ensembling models and DL models with embedding layers tend to outperform other typical models.

\noindent \textit{(4) Both feature abstraction and models' defensive mechanism contribute to detectors' robustness.}
Our analysis indicates that feature abstraction can help ML models capture more robust malicious patterns~\cite{bhagoji2018enhancing}.
For example, MamaDroid~\cite{mariconti2016mamadroid} abstracts program graphs at the family and package levels, which enhances its robustness against adversarial attacks.
In addition, strengthening the defensive capabilities of ML models can further improve detector reliability and stability~\cite{papernot2016distillation}.
As an illustration, RAMDA~\cite{li2021robust} employs an autoencoder to reconstruct input features, learning compact and robust app representations that mitigate the impact of adversarial perturbations.

\noindent \textit{(5) Detection effectiveness does not positively correlate with efficiency.}
Through a combined analysis of effectiveness and efficiency, we observe that effectiveness and efficiency do not always positively correlate.
A detector demanding more resources does not necessarily deliver enhanced results.
For example, Drebin~\cite{arp2014drebin} can achieve competitive detection performance with relatively low resource consumption.

\bulletpoint{Recommendations}
Our findings reveal that the effectiveness of ML-based Android malware detectors is closely tied to the quantity and quality of malware semantics extracted from APK features, which describe the malicious behaviors of apps.
To design more effective and robust detectors, we recommend the following strategies:

\noindent \textit{(1) Quantify malware semantics.}
Rather than indiscriminately incorporating additional features to enhance detection performance, a more principled approach is to design metrics that explicitly evaluate both the quantity and quality of malware semantics captured by features, as well as the degree of redundancy among them.
Such metrics could measure properties such as semantic coverage, discriminative power, and feature overlap across different behavior representations.
By systematically quantifying these aspects, researchers can better identify features that are not only informative but also complementary, enabling more effective feature selection and combination strategies. 
Ultimately, this direction can lead to malware detectors that achieve stronger performance with fewer, semantically richer features, while also improving efficiency and robustness in real-world deployment scenarios.

\noindent \textit{(2) Investigate feature combination and abstraction.}
Given the diverse features available to describe app behaviors, a promising research direction is to explore effective strategies for combining these features to amplify their collective ability to capture malware semantics. 
Additionally, feature abstraction has demonstrated potential in enhancing the robustness of detectors. 
Future work could focus on developing techniques to abstract features in ways that improve detectors' resilience to real-world challenges while maintaining or even enhancing detection performance.

\noindent \textit{(3) Mine invariant malware semantics.}
Noticeable challenges in ML-based Android malware detection include vulnerability to malware evolution and susceptibility to adversarial attacks, both of which can significantly degrade detection performance in real-world deployments.
To mitigate these challenges, future research should focus on identifying and modeling invariant malware semantics—high-level behavioral characteristics that remain stable across malware variants, transformations, and obfuscation techniques.
Such semantics may capture fundamental malicious intents, such as unauthorized data access or covert communication patterns, rather than surface-level syntactic features.
By grounding detection models in these invariant semantics, it becomes possible to build detectors that are more resilient to code evolution, packing, and adversarial manipulation.

\noindent \textit{(4) Align model complexity with malware semantics.}
Once the relevant features and their roles in capturing malware semantics are identified, an equally critical step is selecting models that can effectively leverage this semantic information.
Model complexity should be carefully aligned with the richness and structure of the extracted semantics to achieve an optimal balance between detection effectiveness and computational efficiency.
Overly complex models may fail to provide additional benefits when the input features lack sufficient semantic depth, while overly simplistic models may be incapable of exploiting rich behavioral representations.
Future detector design should therefore emphasize feature-model co-design, ensuring that model expressiveness is commensurate with the semantic information available, particularly in resource-constrained or real-world deployment scenarios.

\section{Discussion}
\label{sec:discussion}

\begin{figure}[t]
    \centering
    \includegraphics[width=0.76\linewidth]{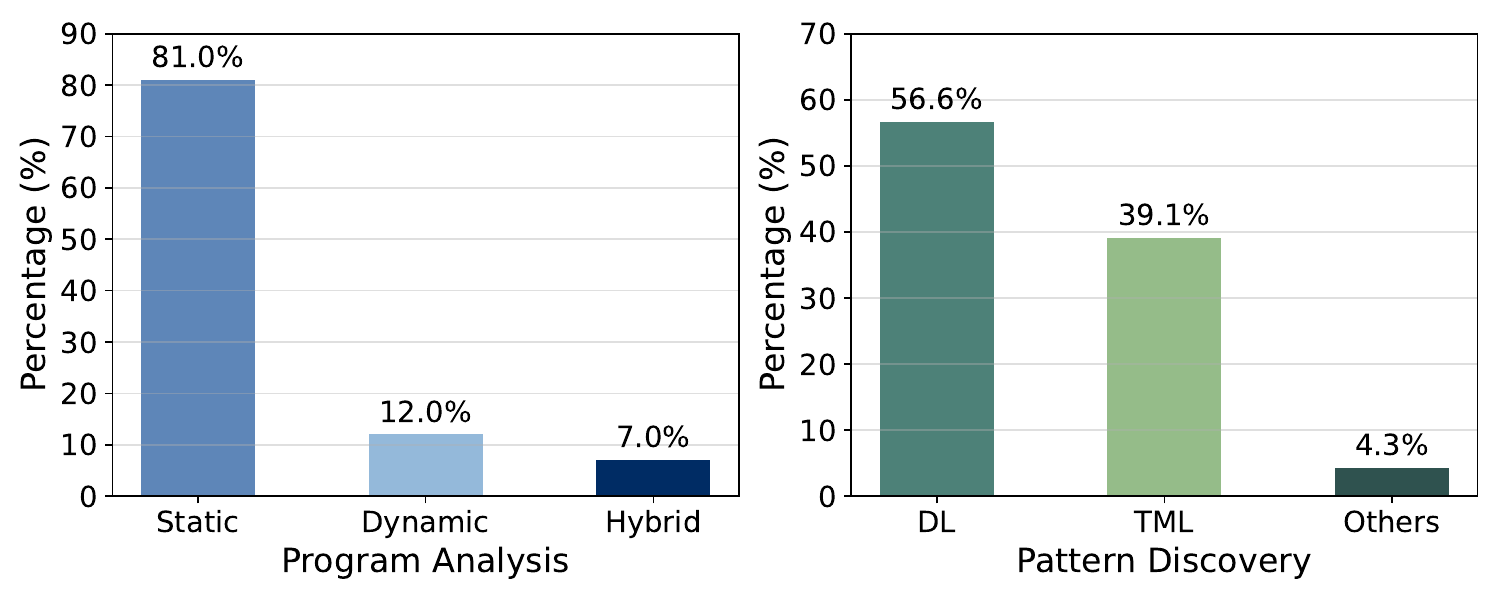}
    \caption{The overall distribution of investigated approaches in Android malware detection.}
    \label{fig:paper_distribution}
    \vspace{-0.3cm}
\end{figure}

\bulletpoint{Systematic investigation}
To better understand the efforts dedicated to Android malware detection, we conduct a systematic literature review since 2011.
Aiming for the inclusion of a broad range of papers, we follow the search strategies used in~\cite{liu2022deep,liu2020review}.
Specifically, we perform searches in key digital libraries, such as the ACM Digital Library and IEEE Xplore, using specific keywords, including \texttt{android malware detection}, \texttt{android analysis}, and \texttt{android malware}.
Then, a careful screening of titles and introductions is followed to selectively exclude studies unrelated to our research topic.
This meticulous process ultimately leads to the identification of 258 related papers.

We subsequently categorize these papers based on the program analysis and pattern discovery techniques they employ.
Figure~\ref{fig:paper_distribution} shows the distribution of these techniques.
Our analysis reveals that \textit{static analysis} is the predominant technique utilized to extract features from apps, followed by \textit{dynamic} and \textit{hybrid analysis}.
For pattern discovery, \textit{machine learning}, including both traditional machine learning (TML) and deep learning (DL) models, are extensively used to identify malicious patterns.
Particularly notable is the significant increase in the utilization of static feature extraction and ML-based techniques in recent years. 
This evolving trend highlights the importance of our study, seeking to provide an in-depth understanding of the contemporary landscape in ML-based Android malware detection.

\bulletpoint{Dataset considerations}
Including mobile apps from multiple markets is essential for improving the representativeness of an Android malware dataset.
To incorporate diverse app sources, we select AndroZoo~\cite{allix2016androzoo} as the primary data source, following common practice in prior studies~\cite{chen2023continuous,pendlebury2019tesseract}. 
AndroZoo aggregates apps collected from multiple distribution channels, including Google Play as well as third-party markets such as Anzhi and AppChina, enabling the construction of a dataset that spans different app ecosystems rather than relying on a single market.

However, we acknowledge certain limitations of using AndroZoo as the primary malware source. 
Although AndroZoo integrates samples from various sources (e.g., VirusShare, AppChina, and Anzhi), a substantial portion of its apps originate from Google Play. 
As a result, some malware samples in AndroZoo may have previously passed Google Play's vetting process at some point in their lifecycle, which may introduce a potential bias. 
Specifically, such samples might not fully capture the characteristics of in-the-wild malware ecosystems, particularly those distributed through less regulated or underground channels.
Beyond this, AndroZoo also does not cover all existing Android markets. 
For example, some prominent regional app stores—such as Huawei AppGallery and other OEM-operated markets—are not included. 
Incorporating apps from these additional sources, as well as alternative malware feeds, could further enhance dataset comprehensiveness and better reflect the diversity of the Android threat landscape.

Nevertheless, during dataset construction, we explicitly ensured that our samples include applications originating from multiple markets within AndroZoo, including Google Play, VirusShare, AppChina, Anzhi, and VirusTotal, rather than being dominated by a single source.
Consequently, our analysis and findings are grounded in behaviors observed across diverse app markets, which mitigates the risk that our conclusions are driven by market-specific biases.
Moreover, different app markets adopt security policies and vetting mechanisms, which can affect both the prevalence and behavioral characteristics of malware.
Although these vetting strategies often share common mechanisms --- such as signature-based scanning and heuristic analysis --- prior work has shown that they are inherently reactive and subject to non-negligible detection delays~\cite{grace2012riskranker,zhou2012dissecting}.
To account for this, we adopt a long time span when constructing our dataset, allowing malicious apps to be labeled post hoc and increasing the likelihood that the collected malware samples are representative and behaviorally diverse.
While acknowledging the dataset's limitations, we believe that our primary dataset is sufficiently diverse and representative to support a meaningful and realistic evaluation of ML-based Android malware detectors.

We assess the performance of the selected Android malware detectors using two datasets collected over different time periods: a primary dataset spanning 2010-2020 and an additional dataset from 2021-2024. The additional dataset is used to evaluate effectiveness, allowing us to examine whether detection performance trends remain consistent across temporally distinct Android ecosystems.

For the effectiveness evaluation, we observe that detection trends are consistent across both datasets, with similar performance rankings among the evaluated approaches. For example, HinDroid, Kim et al., and HomDroid achieve top performance in both time periods. This consistency indicates that our key findings regarding detection effectiveness are robust and generalizable across different time periods.
We further examine performance variations across datasets from different time periods in Section~\ref{effectiveness} (\textit{Detector stableness of different time period dataset}). The results show that the effectiveness of these detectors remains relatively stable across the two periods, with only minor fluctuations in performance metrics. This observation further supports the conclusion that our findings derived from the primary dataset are applicable to more recent Android ecosystems as well.
For the robustness evaluation, the results are conceptually insensitive to the specific time period of the dataset. In malware evolution experiments, the key requirement is that training samples precede testing samples temporally. For obfuscation and adversarial evaluations, robustness is assessed by applying controlled transformations to testing apps, which does not depend on the absolute collection years of the dataset.
For the efficiency evaluation, we measure performance using the primary dataset while explicitly accounting for temporal changes in app characteristics. The decade-long span of this dataset is sufficient to capture substantial changes in app complexity and scale, enabling a meaningful assessment of efficiency trends.

Overall, while the reliance on AndroZoo introduces inherent limitations in fully capturing the in-the-wild malware landscape, our dataset provides a large-scale, systematically curated, and widely adopted benchmark. 
Combined with cross-temporal evaluation and complementary robustness and efficiency analyses, it supports a meaningful and reproducible assessment of ML-based Android malware detectors.

\bulletpoint{App packing and its impact}
To protect applications from reverse engineering and tampering, many developers employ app packing techniques that obfuscate the original code structure~\cite{dong2022did,duan2018things}.
Such techniques can substantially degrade the effectiveness of static analysis-based malware detectors, as they hinder the extraction of meaningful features from packed apps.
In this study, to systematically examine the state of ML-based Android malware detection, we focus primarily on unpacked apps, which is consistent with the scope of most prior work~\cite{chen2023continuous,wu2021homdroid}.
During dataset construction, we identify an app as packed if it employs any known packing tools (\textit{e.g.,} Qihoo 360) and exclude these apps to ensure reliable static feature extraction.
We acknowledge that packing techniques can increase static feature extraction failure rates, negatively impacting the performance of static analysis-based detectors.
To mitigate this limitation in practical deployments, future detectors may benefit from integrating static and dynamic analysis techniques, enabling more robust feature extraction in the presence of packing and obfuscation.
In this work, our objective is to evaluate and understand the performance characteristics of ML-based Android malware detectors under controlled and comparable conditions.
A systematic investigation of app packing techniques and their impact on malware detection effectiveness is an important and complementary direction, which we leave for future work.

\bulletpoint{Potential of large language models in Android malware analysis}
Large Language Models (LLMs) have demonstrated remarkable capabilities in natural language processing, code understanding, and reasoning tasks.
Building upon the findings and insights from our study, we believe that LLMs hold substantial promise in the domain of Android malware analysis, particularly in the following areas:
(1) Feature space exploration: LLMs can assist in selecting and combining discriminative features to enhance the performance of malware detectors.
By analyzing the impact of different feature sets on app behavior description, LLMs can guide the construction of more effective and concise feature representations.
(2) Semantic understanding of malicious behavior: LLMs can mine high-level malicious semantics directly from code, improving the interpretability of detection systems and offering deeper insight into the behavioral mechanisms of malware.

For example, LLMs can analyze multiple features that describe the same malicious behavior, evaluate their relative impact on detection performance, and identify the most informative feature combinations for comprehensive behavior modeling.
Moreover, LLMs can summarize the semantics of groups of functions within an application to uncover specific malicious operations—such as sensitive information collection, data exfiltration, or command-and-control communication.
These semantic summaries can then be leveraged to enhance app characterization and improve the detection of potentially malicious behaviors.

\bulletpoint{Threats to validity}
\jh{There are two main threats to the validity of our study.}
First, our research mainly focuses on investigating general ML-based Android malware detectors.
That is, we do not include the methods designed to solve a particular challenge like malware evolution.
Specifically, there have been several attempts~\cite{narayanan2016adaptive,zhang2020enhancing,chen2023continuous,xu2019droidevolver} starting to mitigate the challenges we have identified.
For instance, the recent APIGraph~\cite{zhang2020enhancing} identifies semantically similar API calls to enhance detectors' robustness against malware evolution.
Integrating this strategy with popular detectors like Drebin and MamaDroid leads to 5\% - 10\% detection improvements over a one-year malware evolution.
Nevertheless, our findings still hold, as these improvements are insufficient compared to the reduction of around 30\% observed in our study.
It is our aspiration that this study can motivate more researchers to focus on these challenges and develop effective solutions to mitigate them.

Second, inconsistencies might exist between our evaluation and the reported ones due to different settings, such as datasets, metrics, and toolchains.
To mitigate experimental biases and provide a fair comparison, we design a general-purpose framework.
Specifically, we adopt standardized techniques across all tasks, including feature extraction and ML modeling.
We also incorporate many evaluation scenarios, such as training data sizes, malware evolution, and efficiency, to ensure a comprehensive measurement using our crafted dataset.
It is our hope that the framework can facilitate future work in ML-based Android malware detection.

\section{Conclusion}
\label{sec:conclusion}
This paper performs the most extensive systematic study of the ML-based Android malware detection literature with empirical and quantitative analysis.
We identify challenges (\textit{i.e.,} unfair comparisons, unrealistic evaluations, and unclear computational costs) that hinder the systematization in this field.
In response, we design a general-purpose framework for developing ML-based detection approaches and evaluating their effectiveness, robustness, and efficiency.
By experimentally comparing 12 representative approaches, our study paints a holistic view of the state of ML-based Android malware detection and puts forth recommendations to guide future research.
For instance, we find existing detectors are still vulnerable to malware evolution and adversarial attacks and in future work, we should focus on incorporating more \textit{malware semantics} to design more practical detectors.

\section*{Acknowledgments}
We thank Bonan Ruan, Jiawei Li, and Chuqi Zhang for their assistance and the anonymous reviewers for their valuable comments.
This research is partially supported by the National Research Foundation, Singapore, through the National Cybersecurity R\&D Lab at the National University of Singapore under its National Cybersecurity R\&D Programme (Award No. NCR25-NCL P3-0001) and by UK EPSRC Grant no. EP/X015971/2. 
Any opinions, findings and conclusions or recommendations expressed in this material are those of the authors(s) and do not reflect the views of National Research Foundation, Singapore, National Cybersecurity R\&D Lab at the National University of Singapore, and EPSRC.

\clearpage
\bibliographystyle{ACM-Reference-Format}
\bibliography{paper}
\clearpage
\appendix
\section{Appendix}
This section contains additional data and information that can be of interest to the readers but that are not strictly necessary for understanding the main text.

\subsection{Feature Database}
\label{sub:feature_db}
To streamline the evaluation of different approaches, we establish a feature database to maintain commonly used features in Android malware detection.
Below, we outline the main features incorporated into the database, categorizing them for ease of reference.
\begin{itemize}[leftmargin=10pt]
    \setlength\itemsep{-1pt}
    \item \textit{Manifest}: this category contains features sourced from the \textit{AndroidManifest.xml} file, such as hardware components, permissions, intents, and app components.
    \item \textit{DisassembledCode}: this category includes the disassembled data extracted from the DEX file, such as the opcode, operands, and code strings.
    \item \textit{ProgramGraph}: this is mainly used to store the program graph of the APK file, including the nodes and edges.
    \item \textit{SharedLibrary}: this category contains the information of the shared libraries used by the APK file.
    \item \textit{Others}: this category includes other features that are not covered by the above categories.
\end{itemize}
Given the structure of the feature database, adding new features becomes straightforward, facilitating the adaptive evaluation of diverse approaches.
For instance, we can easily implement an add-on feature extractor and store the derived features in the database, waiting for use by the preprocessor.

\begin{table*}[t]
    \centering
    \caption{
    An overview of the 12 representative approaches we selected from major venues in security, software engineering, and machine learning. 
    \ymark \ indicates that the APK file (\textit{e.g.}, \textit{AndroidManifest.xml} for Manifest) is taken as input in Android malware detection, while \nmark \ is the opposite.
    \cmark \ and \xmark \ indicate whether feature engineering is handcrafted using domain knowledge or learned using representation learning.
    The effectiveness is based on the results reported in the original papers.}
    \vspace{-0.1cm}
    \begin{adjustbox}{width=\linewidth}
    \begin{tabular}{@{}c|cc|cccc|cc|rr|rrr@{}}
    \toprule
    \multirow{2}{*}{\raisebox{-0.3cm}{\textbf{\begin{tabular}[c]{@{}c@{}}Selected \\ Approach\end{tabular}}}} & \multicolumn{2}{c|}{\textbf{Publication}} & \multicolumn{4}{c|}{\textbf{Input from APK}} & \multicolumn{2}{c|}{\textbf{Feature Engineering}} & \multicolumn{2}{c|}{\textbf{Dataset}} & \multicolumn{3}{c}{\textbf{Effectiveness}} \\ \cmidrule(lr){2-3} \cmidrule(lr){4-7} \cmidrule(lr){8-9} \cmidrule(lr){10-11} \cmidrule(l){12-14}
     & \textbf{Venue} & \textbf{Year} & \textbf{Manifest} & \textbf{Dex} & \textbf{Resource} & \textbf{Library} & \textbf{Handcrafted} & \textbf{Learned} & \textbf{Malware} & \textbf{Goodware} & \textbf{TPR} & \textbf{FPR} & \textbf{F1} \\ \midrule
    Drebin\cite{arp2014drebin} & NDSS & 2014 & \ymark & \ymark & \nmark & \nmark & \cmark & \xmark & 5,560 & 123,453 & 94\% & \textbf{1\%} & 87\% \\ \midrule
    MamaDroid\cite{mariconti2016mamadroid} & NDSS & 2017 & \nmark & \ymark & \nmark & \nmark & \cmark & \xmark & 35,493 & 8,447 & 97\% & 2\% & 96\% \\ \midrule
    Mclaughlin et al.\cite{mclaughlin2017deep} & CODASPY & 2017 & \nmark & \ymark & \nmark & \nmark & \xmark& \cmark & 13,637 & 13,758 & 95\% & \textbf{1\%} & 97\% \\ \midrule
    HinDroid\cite{hou2017hindroid} & KDD & 2017 & \nmark & \ymark & \nmark & \nmark & \cmark & \xmark & 1,216 & 1,118 & \textbf{99\%} & 2\% & \textbf{99\%} \\ \midrule
    DeepRefiner\cite{xu2018deeprefiner} & EuroS\&P & 2018 & \ymark & \ymark & \ymark & \nmark & \xmark & \cmark & 62,915 & 47,525 & 98\% & 2\% & 98\% \\ \midrule
    Kim et al.\cite{kim2018multimodal} & TIFS & 2018 & \ymark & \ymark & \nmark & \ymark & \cmark & \cmark & 21,260 & 20,000 & \textbf{99\%} & \textbf{1\%} & \textbf{99\%} \\ \midrule
    MalScan\cite{wu2019malscan} & ASE & 2019 & \nmark & \ymark & \nmark & \nmark & \cmark & \xmark & 15,430 & 15,285 & --- & --- & 98\% \\ \midrule
    SDAC\cite{xu2020sdac} & TDSC & 2020 & \nmark & \ymark & \nmark & \nmark & \cmark & \cmark & 34,497 & 35,645 & 98\% & \textbf{1\%} & \textbf{99\%}\\ \midrule
    HomDroid\cite{wu2021homdroid} & ISSTA & 2021 & \nmark & \ymark & \nmark & \nmark & \cmark & \xmark & 3,358 & 4,840 & 97\% & 4\% & 95\% \\ \midrule
    Xmal\cite{wu2021android} & TOSEM & 2021 & \ymark & \ymark & \nmark & \nmark & \cmark & \xmark & 15,570 & 20,120 & 98\% & 2\% & 98\% \\ \midrule
    RAMDA\cite{li2021robust} & WWW & 2021 & \ymark & \ymark & \nmark & \nmark & \cmark & \xmark & 21,621 & 36,862 & 93\% & \textbf{1\%} & 90\% \\ \midrule
    MSDroid\cite{he2022msdroid} & TDSC & 2022 & \nmark & \ymark & \nmark & \nmark & \cmark & \xmark & 30,210 & 51,580 & 97\% & \textbf{1\%} & 97\% \\ \bottomrule
    \end{tabular}
    \end{adjustbox}
    \centering
    \begin{tablenotes}
    \footnotesize
    \item \ \ TPR, FPR, and F1 refer to true positive rate, false positive rate, and F1-score. 
    As MalScan is only evaluated on F1, its TPR and FPR are not available.
    \end{tablenotes}
    \label{tab:literature}
    \vspace{-0.3cm}
\end{table*}

\subsection{Reproduction of Selected Approaches}
\label{sub:approach_overview}
Table~\ref{tab:literature} offers a summary of the 12 representative approaches we have selected from leading publications in security~\cite{arp2014drebin,mariconti2016mamadroid,mclaughlin2017deep,xu2018deeprefiner,kim2018multimodal,xu2020sdac,he2022msdroid}, software engineering~\cite{wu2019malscan,wu2021homdroid,wu2021android}, and machine learning~\cite{hou2017hindroid,li2021robust}.
Within this table, we outline the publication details, input from APK, feature engineering style, dataset statistics, and their original effectiveness.
To better support subsequent research and foster replicability, we provide the hyper-parameters of these approaches.

\bulletpoint{Drebin}
We replicate Drebin~\cite{arp2014drebin} utilizing a linear SVM with $C=1$, and set max number of iterations to $1000$.

\bulletpoint{MamaDroid}
MamaDroid~\cite{mariconti2016mamadroid} has two abstract mode \textit{i.e.}, package and family.
In our experiments, we chose the family mode, as it is more efficient in terms of time and memory.
For the RF classifier, we employ the default hyper-parameters provided by scikit-learn.

\bulletpoint{Mclaughlin et al}
Mclaughlin et al.~\cite{mclaughlin2017deep} is implemented with a CNN with one single convolutional layer, followed by a max-pooling layer and a fully connected layer.
The convolutional layer is configured with $32$ filters and a kernel size of $225*7$.
The fully connected layer has $16$ neurons.
For the evaluation process, a learning rate of $0.01$ and a batch size of $32$ are employed.
Additionally, inputs that exceed a length of $600000$ are truncated to $60000$, and those that are less than $600000$ are padded with zeros.

\bulletpoint{HinDroid}
In implementing HinDroid~\cite{hou2017hindroid}, various meta-path combinations have been tested. 
$AA^{T}$ yields remarkable performance in our experiments, and hence, is chosen as the meta-path.
For multi-kernel learning, we utilize the $p-norm$ multi-kernel learning framework as released by~\cite{sun2010multiple}, applying the default settings for the hyper-parameters.

\bulletpoint{DeepRefiner}
As discussed in Section~\ref{sub:selected_approaches}, DeepRefiner~\cite{xu2018deeprefiner} has two detection layers, and the second layer shows strong detection capability.
In our experiments, we replicate the second LSTM-based detection layer.
The model configuration is as follows: an embedding size of $16$ for word embeddings (bytecode instructions), a two-layer LSTM model with an input size of $16$, and a hidden size of $64$.
Batch size is set as $32$, and the learning rate is $0.001$.
The input sequence has a maximum length of $50,000$.
Inputs exceeding this length are truncated, while shorter inputs are padded with zeros.

\bulletpoint{Kim et al}
Kim et al.~\cite{kim2018multimodal} initially use a $5$ separated MLPs to process features from five various modalities, each with the same configuration.
Subsequently, a new MLP is introduced to integrate the learned features from the previous MLPs.
The initial five MLPs have layer configurations of $5000, 2500$, and $1000$ neurons.
The last MLP is structured with layers containing $1000, 500, 100$, and $10$ neurons, respectively.
A learning rate of $0.001$ and a batch size of $32$ are used during the training process.

\bulletpoint{MalScan}
The algorithm~\cite{wu2019malscan} is replicated with a KNN classifier, with the number of neighbors set to $3$.

\bulletpoint{SDAC}
SDAC~\cite{xu2020sdac} initially employs Word2Vec to encode API calls into vectors. Following this, K-means clustering is applied, and then these cluster centers are used as anchors to encode features.
In the classification phase, for simplicity, we use a linear SVM instead of multi-voting SVMs.
Additionally, due to K-means's randomness, the algorithm is executed $5$ times, and the average results are computed.
The size of Word2Vec embedding is set as $10$.
The chosen number of clusters is $1000$, and the SVM is configured using default parameters, allowing for $5000$ iterations.

\bulletpoint{HomDroid}
The approach~\cite{wu2021homdroid} is executed using a KNN classifier, with the number of neighbors set to $1$.

\bulletpoint{Xmal}
Xmal~\cite{wu2021android} is implemented using a 3-layer MLP with a hidden size of $64$.
An attention layer is also incorporated, utilizing an MLP with a hidden size of $158$.
The learning rate is set as $0.001$, and the batch size is $20$.

\bulletpoint{RAMDA}
RAMDA~\cite{li2021robust}'s architecture consists of two parts: autoencoder and classifier.
The encoder and decoder each consist of a 3-layer MLP with a hidden size of $600$.
The classifier is a 4-layer MLP with a hidden size of $600$.
During the training process, the autoencoder is trained first, and then the classifier is trained.
Configuration parameters are established with a learning rate of $0.001$, a batch size of $64$, and epochs set at $20$.
A pre-defined reconstruction loss of $30$ is used.
To constrain the loss, the positive weights are defined as $\lambda_1 = 10, \lambda_2 = 1, \lambda_3 = 10$.

\bulletpoint{MSDroid}
MSDroid~\cite{he2022msdroid} is implemented using a 3-layer GNN with a hidden size of $512$.
Subsequent to this, a 2-layer fully connected network with a hidden size of $512$ is used to classify the APKs.
The model's training parameters are set with a learning rate of $0.01$ and a batch size of $64$.

\begin{table*}[t]
    \centering
    \renewcommand{\arraystretch}{1.1}
    \caption{The AUT(TPR, N) and AUT(FRP, N) of the selected approaches over varying time decays. In this analysis, we examine the shifts during various malware evolution periods: 3, 6, 9, 12, 15, 18, 21, and 24 months.}
    \vspace{-0.1cm}
    \begin{adjustbox}{width=\textwidth}
        \begin{tabular}{@{}c|c|c|c|c|c|c|c|c|c@{}}
            \toprule
            \textbf{Approach}         & \textbf{AUT(TPR,0)} & \textbf{AUT(TPR,3)} & \textbf{AUT(TPR,6)} & \textbf{AUT(TPR,9)} & \textbf{AUT(TPR,12)} & \textbf{AUT(TPR,15)} & \textbf{AUT(TPR,18)} & \textbf{AUT(TPR,21)} & \textbf{AUT(TPR,24)} \\ \midrule
            Drebin            & 0.803               & 0.722(-10.1\%)      & 0.670(-16.5\%)      & 0.601(-25.1\%)      & 0.564(-29.8\%)       & 0.530(-34.0\%)       & 0.508(-36.7\%)       & 0.484(-39.7\%)       & 0.466(-41.9\%)       \\ \midrule
            MamaDroid         & 0.712               & 0.496(-30.3\%)      & 0.429(-39.7\%)      & 0.365(-48.7\%)      & 0.322(-54.7\%)       & 0.308(-56.7\%)       & 0.282(-60.4\%)       & 0.253(-64.5\%)       & 0.233(-67.2\%)       \\ \midrule
            Mclaughlin et al. & 0.724               & 0.583(-19.6\%)      & 0.526(-27.3\%)      & 0.461(-36.4\%)      & 0.410(-43.4\%)       & 0.378(-47.9\%)       & 0.354(-51.2\%)       & 0.327(-54.9\%)       & 0.303(-58.1\%)       \\ \midrule
            HinDroid          & 0.831               & 0.786(-5.5\%)       & 0.773(-7.0\%)       & 0.729(-12.3\%)      & 0.687(-17.4\%)       & 0.684(-17.8\%)       & 0.674(-19.0\%)       & 0.662(-20.3\%)       & 0.655(-21.2\%)       \\ \midrule
            DeepRefiner       & 0.774               & 0.590(-23.7\%)      & 0.545(-29.6\%)      & 0.478(-38.3\%)      & 0.427(-44.8\%)       & 0.413(-46.6\%)       & 0.388(-49.9\%)       & 0.362(-53.2\%)       & 0.339(-56.2\%)       \\ \midrule
            Kim et al.        & 0.845               & 0.760(-10.0\%)      & 0.745(-11.8\%)      & 0.676(-20.0\%)      & 0.619(-26.7\%)       & 0.583(-31.0\%)       & 0.557(-34.0\%)       & 0.524(-38.0\%)       & 0.49(-42.0\%)        \\ \midrule
            MalScan           & 0.800               & 0.685(-14.4\%)      & 0.650(-18.8\%)      & 0.584(-27.0\%)      & 0.547(-31.6\%)       & 0.519(-35.2\%)       & 0.494(-38.3\%)       & 0.465(-41.8\%)       & 0.439(-45.1\%)       \\ \midrule
            SDAC              & 0.734               & 0.616(-16.1\%)      & 0.554(-24.5\%)      & 0.495(-32.6\%)      & 0.455(-38.0\%)       & 0.439(-40.2\%)       & 0.416(-43.4\%)       & 0.391(-46.8\%)       & 0.369(-49.7\%)       \\ \midrule
            HomDroid          & 0.836               & 0.689(-17.5\%)      & 0.651(-22.1\%)      & 0.596(-28.7\%)      & 0.546(-34.6\%)       & 0.515(-38.4\%)       & 0.486(-41.8\%)       & 0.448(-46.4\%)       & 0.422(-49.5\%)       \\ \midrule
            Xmal              & 0.836               & 0.741(-11.3\%)      & 0.693(-17.1\%)      & 0.617(-26.1\%)      & 0.575(-31.2\%)       & 0.550(-34.2\%)       & 0.523(-37.4\%)       & 0.492(-41.2\%)       & 0.463(-44.6\%)       \\ \midrule
            RAMDA             & 0.876               & 0.814(-7.0\%)       & 0.775(-11.6\%)      & 0.733(-16.3\%)      & 0.689(-21.3\%)       & 0.673(-23.2\%)       & 0.658(-24.8\%)       & 0.637(-27.2\%)       & 0.613(-30.1\%)       \\ \midrule
            MSDroid           & 0.835               & 0.755(-9.6\%)       & 0.732(-12.3\%)      & 0.673(-19.4\%)      & 0.635(-24.0\%)       & 0.632(-24.3\%)       & 0.604(-27.6\%)       & 0.569(-31.9\%)       & 0.544(-34.8\%)       \\ \midrule \midrule

            \textbf{Approach}         & \textbf{AUT(FPR,0)} & \textbf{AUT(FPR,3)} & \textbf{AUT(FPR,6)} & \textbf{AUT(FPR,9)} & \textbf{AUT(FPR,12)} & \textbf{AUT(FPR,15)} & \textbf{AUT(FPR,18)} & \textbf{AUT(FPR,21)} & \textbf{AUT(FPR,24)} \\ \midrule
            Drebin            & 0.014               & 0.02(+42.2\%)       & 0.023(+68.2\%)      & 0.024(+71.1\%)      & 0.026(+85.6\%)       & 0.029(+110.1\%)      & 0.030(+115.2\%)      & 0.031(+126.7\%)      & 0.033(+141.2\%)      \\ \midrule
            MamaDroid         & 0.009               & 0.010(+17.9\%)      & 0.011(+28.4\%)      & 0.010(+20.2\%)      & 0.011(+31.9\%)       & 0.013(+50.5\%)       & 0.013(+54.0\%)       & 0.014(+62.2\%)       & 0.014(+59.9\%)       \\ \midrule
            Mclaughlin et al. & 0.013               & 0.010(-19.1\%)      & 0.012(-7.5\%)       & 0.013(+1.9\%)       & 0.015(+14.3\%)       & 0.016(+23.6\%)       & 0.016(+24.4\%)       & 0.016(+22.1\%)       & 0.016(+20.5\%)       \\ \midrule
            HinDroid          & 0.016               & 0.262(+1491.6\%)    & 0.266(+1516.6\%)    & 0.271(+1548.8\%)    & 0.279(+1599.3\%)     & 0.319(+1839.1\%)     & 0.321(+1853.7\%)     & 0.324(+1873.8\%)     & 0.327(+1890.3\%)     \\ \midrule
            DeepRefiner       & 0.017               & 0.019(+9.3\%)       & 0.018(+5.9\%)       & 0.019(+7.6\%)       & 0.021(+20.3\%)       & 0.022(+27.2\%)       & 0.022(+28.9\%)       & 0.023(+31.2\%)       & 0.024(+38.7\%)       \\ \midrule
            Kim et al.        & 0.014               & 0.017(+19.7\%)      & 0.018(+29.8\%)      & 0.018(+31.9\%)      & 0.02(+43.5\%)        & 0.024(+75.9\%)       & 0.026(+84.6\%)       & 0.027(+91.8\%)       & 0.027(+95.4\%)       \\ \midrule
            MalScan           & 0.025               & 0.029(+17.1\%)      & 0.029(+18.7\%)      & 0.029(+17.9\%)      & 0.031(+24.8\%)       & 0.031(+26.1\%)       & 0.033(+33.0\%)       & 0.035(+40.7\%)       & 0.037(+50.1\%)       \\ \midrule
            SDAC              & 0.024               & 0.042(+72.1\%)      & 0.050(+104.4\%)     & 0.054(+121.6\%)     & 0.060(+146.5\%)      & 0.064(+159.6\%)      & 0.066(+169.9\%)      & 0.068(+176\%)        & 0.072(+192.3\%)      \\ \midrule
            HomDroid          & 0.014               & 0.023(+63.2\%)      & 0.024(+66.0\%)      & 0.023(+59.0\%)      & 0.024(+66.0\%)       & 0.026(+78.5\%)       & 0.028(+94.6\%)       & 0.031(+112.7\%)      & 0.032(+123.8\%)      \\ \midrule
            Xmal              & 0.023               & 0.025(+11.6\%)      & 0.028(+23.5\%)      & 0.027(+20.4\%)      & 0.027(+18.2\%)       & 0.028(+21.7\%)       & 0.028(+25.3\%)       & 0.029(+28.4\%)       & 0.030(+30.6\%)       \\ \midrule
            RAMDA             & 0.054               & 0.061(+14.0\%)      & 0.069(+27.6\%)      & 0.075(+38.7\%)      & 0.081(+50.0\%)       & 0.09(+67.1\%)        & 0.093(+73.4\%)       & 0.099(+84.2\%)       & 0.104(+93.9\%)       \\ \midrule
            MSDroid           & 0.072               & 0.078(+8.4\%)       & 0.080(+11.7\%)      & 0.082(+14.5\%)      & 0.085(+18.4\%)       & 0.093(+29.2\%)       & 0.098(+37.3\%)       & 0.106(+47.4\%)       & 0.111(+55.1\%)       \\ \bottomrule
            \end{tabular}
    \end{adjustbox}
    \label{tab:tpr_time_decay}
    \vspace{-0.3cm}  
\end{table*}

\begin{table}[t]
    \centering
    \renewcommand{\arraystretch}{0.6}
    \caption{The effectiveness of the selected approaches using different-sized training data.}
    \vspace{-0.1cm}
    \begin{adjustbox}{width=0.54\linewidth}
    \begin{tabular}{@{}c|c|c|c@{}}
    \toprule
    \diagbox[innerleftsep=-0.1em, innerrightsep=-0.2em, height=2.6em,width=7em]{\textbf{Approach}}{\textbf{Data Size}} & \textbf{100\%} & \textbf{50\%}& \textbf{10\%}\\ \midrule
    Drebin& 0.722 & 0.714 (-1.2\%) & 0.643 (-10.9\%) \\ \midrule
    MamaDroid& 0.661 & 0.623 (-5.7\%) & 0.452 (-31.5\%) \\ \midrule
    Mclaughlin et al.& 0.714 & 0.627 (-12.2\%) & 0.140 (-80.4\%) \\ \midrule
    HinDroid & 0.731 & 0.716 (-2.0\%) & 0.618 (-15.5\%) \\ \midrule
    DeepRefiner& 0.657 & 0.577 (-12.1\%) & 0.327 (-50.2\%) \\ \midrule
    Kim et al. & 0.782 & 0.742 (-5.1\%) & 0.611 (-21.8\%) \\ \midrule
    MalScan& 0.684 & 0.653 (-4.6\%) & 0.526 (-23.1\%) \\ \midrule
    SDAC & 0.522 & 0.496 (-5.0\%) & 0.474 (-9.1\%) \\ \midrule
    HomDroid& 0.734 & 0.675 (-8.0\%) & 0.586 (-20.1\%) \\ \midrule
    Xmal& 0.698 & 0.668 (-4.2\%) & 0.613 (-12.2\%) \\ \midrule
    RAMDA& 0.636 & 0.614 (-3.4\%) & 0.547 (-14.0\%) \\ \midrule
    MSDroid& 0.648 & 0.563 (-13.0\%) & 0.424 (-34.6\%) \\ \bottomrule
    \end{tabular}
    \end{adjustbox}
    \label{tab:res_training_size}
    \vspace{-0.3cm}
\end{table}

\subsection{AUT, ASR, and APR}
\label{sub:aut}
In assessing a classifier's resilience against temporal degradation, we employ the Area Under Time (AUT) metric as suggested by~\cite{pendlebury2019tesseract}.
AUT is formally given by:
$$
AUT(f, N) = \frac{1}{N-1}\sum_{k=1}^{N-1}\frac{f(k+1)+f(k)}{2},
$$
where $f$ represents the chosen performance metric (such as F1 score or True Positive Rate) and $N$ denotes the count of malware evolution period. 
AUT values range between $(0,1)$, where 1 indicates the classifier retains consistent performance across time.

To evaluate a classifier's robustness against adversarial attacks, we use two primary metrics: the attack success rate (ASR) and average perturbation ratio (APR) metrics utilized by~\cite{li2023black}.
They are mathematically defined as:
$$
ASR = \frac{N_{success}}{N_{total}}, \ APR = \frac{F_{modified}}{F_{total}}.
$$
Here, $N_{success}$ refers to the number of adversarial examples that successfully deceive the classifier.
$N_{total}$ represents the entire set of adversarial samples.
Meanwhile, $F_{modified}$ is the number of input features changed by the adversarial intervention, and $F_{total}$ signifies the total number of features in the input.
The ASR values fall within the interval $(0,1)$; a value of 1 suggests the classifier is entirely susceptible to adversarial attacks.
Similarly, APR values lie within $(0,1)$.
A higher APR indicates that evading the classifier becomes increasingly challenging.

\subsection{Additional Results}
\label{sec:additional_results}
Table~\ref{tab:res_training_size} presents the detailed results of training data size analysis.
In Sec.~\ref{effectiveness}, we normalize the results of the selected approaches based on the use of 100\% training data size for clarity.
For instance, the normalized result for Drebin, using 50\% of the training data, is computed as $\frac{0.714}{0.722} = 0.988$.
These results highlight the significant impact that the volume of data has on the effectiveness of the selected approaches.

For the analysis of malware evolution, the results of AUT(TPR, N) and AUT(FPR, N) are reported in Table~\ref{tab:tpr_time_decay} (N is also set as 0, 3, 6, 9, 12, 15, 18, 21, and 24 months).
We observe that as malware evolution time increases, there is a consistent decrease in AUT(TPR, N) for the selected approaches.
This trend suggests that as malware evolves, more malware samples are misclassified as benign.
Simultaneously, an increase in AUT(FPR, N) is noted, indicating that more benign samples are misclassified as malware.
Overall, the performance of malware detection approaches degrades with the increase of malware evolution time.

\end{document}